%% file: main.tex
\documentclass[a4paper,12pt,oneside,sfdefaults=false]{scrartcl} 
\usepackage{float} 

\usepackage{ntheorem}
\theoremseparator{:}
\newtheorem{hyp}{Hypothesis}
\usepackage[backend=biber,style=apa,autocite=inline]{biblatex}
\DeclareLanguageMapping{english}{english-apa}
\input{preamble}
\graphicspath{{Figure/}}
\usepackage{threeparttablex}
\usepackage{enumitem}
\usepackage{csquotes}
\usepackage{fontawesome5} 
\usepackage{listings}
\usepackage{lscape}
\lstdefinestyle{RmdStyle}{
    basicstyle=\ttfamily\small,
    numbers=left,
    numberstyle=\tiny\color{gray},
    showstringspaces=false,
    keywordstyle=\color{blue},
    commentstyle=\color{green!50!black},
    stringstyle=\color{red!70!black},
    identifierstyle=\color{black},
    breaklines=true,
    captionpos=b,
    frame=single,
    rulesepcolor=\color{gray},
    tabsize=2
}
\usepackage{makecell}
\usepackage{booktabs}

\newcolumntype{L}{>{\raggedright\arraybackslash}X}

\AtEveryBibitem{\clearfield{isbn}\clearfield{issn}}
\AtEveryBibitem{%
    {
      \iffieldundef{year}
        {}
        {\clearfield{urldate}}
    }
    {}
}
\DeclareSourcemap{
  \maps[datatype=bibtex]{
    \map{
      \step[fieldsource=doi, final]
      \step[fieldset=url, null]
    }
  }
}

\title{\Large Multi-dimensional Bias in Modeling Multi-dimensional Preferences: Evaluating the Ability of Synthetic Agents to Replace Human Participants in Conjoint Experiments\thanks{~~The authors would like to thank Professor Ray Duch, Dr. Jamie Cummins, Dr. Alice Malmberg, and panel participants of IMEBESS 2026 for their valuable feedback. We would also like to thank Professor Ben Ansell, Dr. Noah Bacine, and Anne-Charlotte Gimenez from the Centre for Advanced Social Science Methods (CASSM) at the University of Oxford for their coordination.} \thanks{~~This study was supported by a £1000 research grant from the Department of Politics and International Relations and the Nuffield Centre for Experimental Social Sciences at the University of Oxford.}}

\author{
    \large Ho Ting (Bosco) Hung\thanks{~~Department of Politics and International Relations, University of Oxford} \thanks{~~Oxford Computational Political Science Group} \and
    \large Nachiket Midha\protect \footnotemark[1] \protect\footnotemark[2] \and
    \large Victor Y. Wu\thanks{~~Department of Political Science, Stanford University}  \protect\footnotemark[2] \and
    \large Yiwen Zhang\thanks{~~Department of Government and the Data Science Institute, LSE} \protect\footnotemark[2]
}

\date{\large \today}

\begin{document}

\maketitle

\setcounter{footnote}{0}

\normalsize
\begin{abstract}
\footnotesize
\textbf{Abstract:} Despite growing interest in using LLMs to add robustness or reduce data-collection costs in survey experiments, their efficacy in conjoint design---an increasingly popular method in political science---remains underexplored. This paper addresses that gap by investigating whether synthetic agents can reproduce the multi-dimensional human preference patterns that conjoint is designed to capture. It replicates published conjoint studies and compares the results generated by synthetic agents with original human data along three dimensions: representational correspondence, inferential correspondence, and procedural stability. Our analysis evaluates the alignment of choice distributions as well as the statistical and substantive similarity of estimates, and the results are uneven across these dimensions and studies replicated. This implies that the validity of synthetic participants should be considered claim-dependent and hierarchical. Reproducing a figure or obtaining strong sign agreement is evidence of similar aggregate outputs, but not enough to support replacing human respondents. Our results suggest that the discipline as a whole must first map this innovation's boundaries across various levels before considering synthetic agents a robust substitute for human samples.

\end{abstract}

\doublespacing
\newpage
\justifying

\section{Introduction}
While political choices inherently involve multi-dimensional trade-offs, capturing this complexity has long been challenging. Recent advances, however, alleviate political scientists' struggle to assess competing priorities behind human preferences \autocite{sniderman_advances_2018}. Conjoint experiments unfold opinions and preferences in a multi-dimensional setting, thus enabling political scientists to more effectively study issues with various dimensions like voting and trade policy. As scholars advance experimental innovation, another innovation is under the spotlight: synthetic agents. With the ability to generate human-like texts in response to various inputs, they have drawn scholarly attention for their ability to simulate human preferences \autocite{argyle_out_2023}. 

Despite this enthusiasm for both innovations, there is a surprising lack of studies on whether synthetic agents can be applied specifically in conjoint experiments. This marks a sharp contrast with the increasing number of studies on applying synthetic agents to typical experimental settings \autocites{dillion_can_2023, xie_can_2024, bisbee_synthetic_2024}. However, understanding synthetic agents' effectiveness in various experimental designs is crucial to understanding this innovation's boundary, especially when the quantity of interest in conjoint experiments often differs from that in typical designs.

This paper addresses this key question: Under what conditions can synthetic agents replace human participants in conjoint experiments? By replicating existing conjoint experiments published in major political science journals, we explore how stable and similar the results generated by synthetic agents are relative to those obtained by the original human participants. Our results suggest that the validity of synthetic participants should be understood as claim-dependent and hierarchical, rather than as a binary question of whether they can/cannot replace humans. This is informed by the nuanced finding that synthetic agents often approximate marginal attribute-level distributions and sometimes recover the direction of the estimates, but they perform poorly on full joint profile distributions, individual-level choice alignment, precise effect magnitudes, subgroup heterogeneity, and stability across models.

Moreover, these uneven findings imply that synthetic agents are not reliable substitutes for human samples. Because the studies we replicate are published and widely cited, the aggregate agreement we observe may partly reflect recall of documented findings rather than genuine simulation. While contamination should push synthetic output toward the original human result, we identify that synthetic agents still fail to deliver results aligned with the human sample consistently. Our aggregate results are therefore best read as an upper bound of synthetic performance that signals the need for caution in synthetic agent applications.

This paper's main contributions are as follows: Empirically, as one of the first studies investigating synthetic agents in the context of conjoint experiments, it provides direct evidence on their substitutability in this specific case. By showing how strong performance in typical experimental settings cannot simply be assumed to carry over to the conjoint case, it complements existing research on synthetic agents. Conceptually, it develops a claim-dependent framework for evaluating substitutability, which goes beyond a binary lens examining whether synthetic agents can replace human respondents. Practically, the results point to a validation hierarchy in which the evidence required varies with the intended use, which would help the discipline establish the value and boundaries of this innovation before direct high-stakes deployment.

This paper first reviews relevant existing literature. Second, it discusses the hypotheses. Third, it explains its research design. Fourth, it addresses the empirical analyses. Fifth, it discusses the implications of the findings. Finally, it provides concluding observations.

\section{Literature Review}
\subsection{Conjoint Experiments}
Conjoint experiments have emerged as a key methodological innovation in political science following the seminal work of \textcite{hainmueller_causal_2014}. In such experiments, each attribute is manipulated and can take multiple levels, and then each profile consists of a combination of levels of different attributes, finally forming choice sets presented to the respondents. 

Given their distinct structural architecture, the causal quantity of interest departs from the standard average treatment effects (ATE) used in typical experimental designs: \begin{equation} \text{ATE} = \mathbb{E}[Y(1) - Y(0)] \end{equation} The relatively large number of possible combinations of attribute values makes it unfeasible to estimate the ATEs of particular combinations of attribute values for the full set of attributes included. The expected difference in responses for two different sets of profiles is also less meaningful in practice. Instead, the key quantity of interest is the average marginal component effect (AMCE), which represents the average effect of a particular attribute value of interest against another value of the same attribute while holding equal the joint distribution of the other attributes in the conjoint profile: 
\begin{multline} 
\pi^{(l)}(a_1, a_0) = \sum_{(\mathbf{a}_{-l}, \mathbf{a}') \in \tilde{\mathcal{A}}} \Big\{ \mathbb{E}[Y_{ijk} \mid A_{ijkl} = a_1, \mathbf{A}_{ijk[-l]} = \mathbf{a}_{-l}, \mathbf{A}_{ij[k']} = \mathbf{a}'] \\ 
- \mathbb{E}[Y_{ijk} \mid A_{ijkl} = a_0, \mathbf{A}_{ijk[-l]} = \mathbf{a}_{-l}, \mathbf{A}_{ij[k']} = \mathbf{a}'] \Big\} \cdot \rho(\mathbf{a}_{-l}, \mathbf{a}')
\end{multline}
where $Y_{ijk}$ is the outcome for respondent $i$, task $j$, profile $k$, $A_{ijkl}$ is the level of attribute $l$ in profile $k$, $\mathbf{A}_{ijk[-l]}$ denotes the vector of all other attributes in the same profile, $\mathbf{A}_{ij[k']}$ refers to the attributes of the alternative profile $k'$ in the same task $j$, $\tilde{\mathcal{A}}$ is the support of the joint distribution over attribute combinations, and $\rho(\mathbf{a}_{-l}, \mathbf{a}')$ is the empirical distribution over other attributes used for averaging. 

As there could be heterogeneous treatment effects across various subgroups, another common quantity of interest is the conditional AMCE (cAMCE).\footnote{See \textcite{robinson_how_2024} for a detailed discussion of how to detect heterogeneity in conjoint experiments and several lower-level causal estimands.} It holds a subset of conditioning attributes $\mathbf{A}_{C}$ fixed at a specific joint value $\mathbf{a}_C$. The interaction effects are tested by measuring the marginal effect of attribute $l$ only within the subgroup of profiles defined by $\mathbf{A}_{ijk[C]} = \mathbf{a}_C$, given a particular randomization distribution. cAMCE is defined as: 
\begin{multline} 
\pi^{(l)}_c(a_1, a_0 \mid \mathbf{A}_{ijk[C]} = \mathbf{a}_C) = \sum_{(\mathbf{a}_{R}, \mathbf{a}') \in \tilde{\mathcal{A}}_{R}'} \Big\{ \mathbb{E}[Y_{ijk} \mid A_{ijkl} = a_1, \mathbf{A}_{ijk[C]} = \mathbf{a}_{C}, \mathbf{A}_{ijk[R]} = \mathbf{a}_{R}, \mathbf{A}_{ij[k']} = \mathbf{a}'] \\ 
- \mathbb{E}[Y_{ijk} \mid A_{ijkl} = a_0, \mathbf{A}_{ijk[C]} = \mathbf{a}_{C}, \mathbf{A}_{ijk[R]} = \mathbf{a}_{R}, \mathbf{A}_{ij[k']} = \mathbf{a}'] \Big\} \cdot \rho(\mathbf{a}_{R}, \mathbf{a}') 
\end{multline} 

These two relative estimands can be calculated from another quantity, marginal means (MM).\footnote{See e.g., \textcite{leeper_measuring_2020, abramson_what_2022, treger_changing_2025} for inferential vulnerabilities of AMCEs and why MMs may be more favorable.} It describes the level of favorability toward profiles with a specific feature level, ignoring all other features \autocite{treger_changing_2025}: 
\begin{equation} 
\mu^{(l)}(a_1) = \sum_{(\mathbf{a}_{-l}, \mathbf{a}') \in \tilde{\mathcal{A}}} \Big\{ \mathbb{E}[Y_{ijk} \mid A_{ijkl} = a_1, \mathbf{A}_{ijk[-l]} = \mathbf{a}_{-l}, \mathbf{A}_{ij[k']} = \mathbf{a}'] \Big\} \cdot \rho(\mathbf{a}_{-l}, \mathbf{a}') 
\end{equation}
A combination of MMs allows researchers to compute the choice-level AMCE: \begin{equation} \pi^{(l)}(a_1, a_0) = \mu^{(l)}(a_1) - \mu^{(l)}(a_0) \end{equation} 

\subsection{Synthetic Agents}
Parallel to these advances in experimental design, synthetic agents have gained traction as behavioral proxies for experimental research. Trained using human-generated data, synthetic agents could reflect human preferences to a certain extent and generate fluent responses to inputs in diverse formats \autocites{10.5555/3495724.3495883, horton_large_2023, openai_gpt-4_2023}, as if they interact and reason like humans \autocite{bubeck_sparks_2023}. In political science in particular, scholars like \textcite{wu_large_2023, argyle_out_2023} highlight LLMs' potential in recovering multi-dimensional political preferences, especially as the training data itself already reflects the complicated interplay between political and socio-cultural elements. The ability to simulate aggregate human preferences has been demonstrated by multiple large-scale studies \autocites{park_generative_2024, hewitt_predicting_2024, binz_foundation_2025}. 

Proponents highlight that synthetic agents eliminate traditional survey constraints like fatigue or other pragmatic concerns (e.g., limited attention span, response bias, or habituation) and could process information rapidly, while significantly reducing subject recruitment costs and facilitating validation \autocites{grossmann_ai_2023, argyle_out_2023, dillion_can_2023}. They may also introduce diversity of responses (e.g., race, gender, religion) to facilitate the exploration of intersectionality \autocites{aher_using_2023, argyle_out_2023, byun_dispensing_2023, gerosa_can_2024, shrestha_beyond_2024}. 

Nevertheless, just as many scholars have cast their hope in synthetic agents to revolutionize experimental research, others express substantial concerns \autocite{agnew_illusion_2024}.\footnote{See \textcite{anthis_llm_2025} for a summary of challenges ahead.} At a higher level, the very nature of language models is what drives the criticisms \autocite{bender_climbing_2020, shanahan_talking_2024}. \textcite[617]{bender_dangers_2021} call language models `stochastic parrots' that probabilistically extract information from combined sequences of linguistic features from a vast collection of training data, while not making any reference to meaning. Relatedly, LLMs tend to collapse diverse perspectives into a single modal opinion \autocite{dillion_can_2023}. This is concerning in political science, which cares about preference distributions and not just the average opinion \autocite{bisbee_synthetic_2024}. Even if LLMs can match human participants on overall variability, they may overestimate or underestimate the effect sizes \autocite{wang_not_2024}. These challenges illustrate how the issue of replacement should be approached with nuance. Replacement could mean various things, such as reproducing an aggregate result, recovering the direction of a treatment effect, approximating the distribution of responses, or predicting individual responses. 

Besides, as LLMs are trained on datasets of limited size, concerns about data and algorithmic bias have been raised \autocites{sheng_woman_2019, abid_persistent_2021, ousidhoum_probing_2021, aher_using_2023, dillion_can_2023, durmus_towards_2023, grossmann_ai_2023, hartmann_political_2023, naous_having_2024, santurkar_whose_2023, agnew_illusion_2024}. General AI pitfalls like hallucinations are also often highlighted \autocites{huang_survey_2025, dillion_can_2023, byun_dispensing_2023, agnew_illusion_2024}. This is especially worrying as high-quality data is becoming harder to source \autocite{gao_take_2025}. More technically, value lock-in remains a problem, where a model's static data snapshot fails to adapt to changing real-world events and societal norms \autocites{bender_dangers_2021, weidinger_taxonomy_2022, agnew_illusion_2024, harding_ai_2024}. In short, the validity of synthetic results is not guaranteed.

To address these challenges, one may resort to means like prompt engineering \autocites{kim_interpretability_2018, 10.5555/3495724.3495883, lutz_prompt_2025, anthis_llm_2025}, fine-tuning \autocites{wang_self-instruct_2023, anthis_llm_2025}, or modifying hyperparameters \autocites{radford_improving_2018, fan_hierarchical_2018, hashimoto_unifying_2019, holtzman_curious_2019, gerosa_can_2024, anthis_llm_2025}.\footnote{Note that these solutions often involve trade-offs. For instance, the values of hyperparameters could involve a trade-off between coherence and diversity.} However, even if these solutions might help sometimes, \textcite[337]{11303355} has warned that `a single, one-size-fits-all \textit{homo silicus} is likely a mirage.' Scholars likely require multiple micro-level and domain-specific approaches to ensure the validity of synthetic results, which implies that optimization may not necessarily enhance generalizability. On an aggregate level, \textcite{cummins_threat_2026} offers a novel attempt to understand how these choices (e.g., model choice, hyperparameter setting, scale prompting strategy) may influence the results. Focusing on two social-psychological scales and an interview-style conditioning template tested by \textcite{argyle_out_2023}, he finds that different configuration choices could drive significantly different conclusions.

\subsection{Methodological Intersections}
Despite growing attention to both synthetic agents and conjoint experiments explained above, whether the former can serve the latter remains unexamined. Enthusiasm for using synthetic agents as proxies has focused on simpler experimental settings with different estimands of interest, so demonstrating that they approximate human responses in a basic survey experiment does not extend to other distinct formats. The literature reviewed above further establishes that aggregate resemblance does not by itself establish substitutability. This concern sharpens as tasks complicate, especially when conjoint experiments feature even more complex interactions between contexts and opinions. This motivates a multi-level investigation of this intersection beyond a binary framing of whether synthetic agents can `replace' human participants.

\section{Hypotheses}
We address the following pre-registered hypotheses to evaluate along three dimensions capturing different requirements that synthetic data may be expected to satisfy to mirror human preferences in conjoint experiments.

\subsection{Representational Correspondence}
The first dimension concerns whether synthetic agents reproduce the observed human choice patterns. At the least demanding level, synthetic and human samples may assign similar aggregate weights to individual attribute levels, thereby reproducing the broad relative popularity of the features included in the conjoint design. However, similar marginal distributions do not necessarily imply similar joint distributions. Synthetic agents may reproduce how frequently individual attributes are selected while failing to reproduce how respondents combine those attributes when evaluating complete profiles. An even stronger form of representational correspondence would require synthetic agents to reproduce the choices of the individual human respondents whose demographic characteristics and experimental encounters they mirror. 

\begin{hyp} \label{hyp:distribution}
The distributions of both the overall experimental choices and relative attribute weights between replicating synthetic agents and original human participants in each study will align. 
\end{hyp}

\subsection{Inferential Correspondence}
The second dimension concerns whether synthetic responses support the same substantive and statistical inferences as human responses. Even where the two samples exhibit some degree of distributional resemblance, they may produce different estimates of the quantities of interest. At a basic level, synthetic estimates may recover the direction of human estimates, indicating whether particular attribute levels generally increase or decrease the probability that a profile is selected. A stronger benchmark concerns whether the broader pattern and relative ordering of effects are similar across the two samples. More demanding still is correspondence in the magnitude and uncertainty. A synthetic estimate may have the same sign as its human counterpart while differing sufficiently in size to support a different substantive interpretation. The same requirement applies to heterogeneous effects, which provide an even more demanding test, given their dependence on lower-level variation. 

\begin{hyp} \label{hyp:similarity} 
The estimates from the synthetic agent replications will be similar in direction, magnitude, and significance to those reported in the original human participant studies.
\end{hyp}

\subsection{Procedural Stability}
The third dimension concerns whether synthetic estimates remain stable across reasonable analytic choices in the generation procedure. Correspondence observed under one model or parameter setting may be contingent on a favorable implementation (possibly due to luck or deliberate selection) rather than reflecting a general capacity to reproduce human preferences, similar to the case of $p$-hacking in general data analysis. Such robustness tests are especially important because there is no external basis for identifying one specification as the uniquely appropriate representation of the target population. 

\begin{hyp} \label{hyp:stability}
The estimates from the synthetic agent replication will be robust and stable across various analytic choices.
\end{hyp}

\section{Research Design}
\subsection{Scope}
To test the hypotheses above, we follow \textcite{clayton_correcting_2026} and replicate six studies covering 13 experimental setups listed in Table~\ref{tab:replicated_studies}.\footnote{Note that information about the order/combination of profiles included in each task of each respondent in \textcite{hankinson_when_2018} is missing.} 

\begin{table}[htbp]
\centering
\caption{Selected Studies for Replication}
\footnotesize
\label{tab:replicated_studies}
\begin{tabularx}{\textwidth}{p{6cm} X r}
\toprule
\textbf{Authors} & \textbf{Topic} & \textbf{Tasks $\times$ Sample} \\
\midrule
\textcite{arias_changing_2022} & Climate Migration & 19,440 \\
\textcite{bechtel_mass_2013} & Global Climate Agreements & 34,000 \\
\textcite{hainmueller_hidden_2015} & Immigration & 7,035 \\
\textcite{hankinson_when_2018} & Housing & 15,095 \\
\textcite{ono_contingent_2019} & Candidate Gender Effects & 15,830 \\
\textcite{teele_ties_2018} & Women in Politics & 11,955 \\
\bottomrule
\end{tabularx}
\end{table}

\subsection{Synthetic Sample Generation}
We generate the sample using the following mainstream models at the time of writing: GPT-4o, GPT-4o mini, Llama 3.2 (3B), Llama 3.3 (70B), and Gemini 2.5 Flash. We avoid overly advanced models, which may perform costly and time-consuming logical deduction rather than quickly reflecting human choices. Thus, our evidence speaks to realistic and affordable research practices rather than to a capability ceiling that few applied researchers can afford to reach, especially as conjoint design tends to involve longer input. In addition, as an open-source model, Llama allows better reproducibility.

We use a moderate temperature $T=0.5$ as a baseline specification to balance diversity and predictability. We adopt a one-to-one persona mirroring strategy and benchmark synthetic agents to reflect the same set of demographic information and profile encounters as the human sample. This helps minimize the design variance. Additionally, we perform an exploratory exercise by tweaking $T \in \{0.0, 0.5, 1.0\}$ when testing \textcite{teele_ties_2018, hainmueller_hidden_2015, hankinson_when_2018, ono_contingent_2019}, which are chosen based on the sum of profiles included.

\subsection{Statistical Tests}
\subsubsection{Distribution}
\label{test:distribution}
To evaluate Hypothesis~\ref{hyp:distribution}, we compare the general choice distributions at joint, marginal, and individual levels.\footnote{Per \textcite{hainmueller_causal_2014}, we note that designs with attribute constraints may have different baseline distributions from designs without constrained attributes.} We start with the Wasserstein distance to compare the full joint distribution of chosen profiles \autocite{kantorovich_mathematical_1960}: \begin{equation} W(P_H, P_A) = \inf_{\Gamma \in \Pi(P_H, P_A)} \sum_{i,j} \Gamma_{i,j} d_{i,j} \end{equation} where $\Pi(P_H, P_A)$ is the set of all joint distributions whose marginals are $P_H$ and $P_A$, and $\Gamma_{i,j}$ is the `flow' from profile $i$ in $P_H$ to profile $j$ in $P_A$. The term $d_{i,j}$ represents the entries of a cost matrix $M$, where each $M_{i,j}$ is calculated as the Manhattan distance between the one-hot-encoded attribute profiles $i$ and $j$. This `ground distance' ensures the metric remains informative despite the sparsity of the joint profile space in conjoint designs, as it penalizes mass transport based on the degree of attribute-level dissimilarity. Since $W(P_H, P_A)$ is a descriptive distance metric without an associated sampling distribution, we use non-parametric bootstrapping ($B=2000$) of human data to establish a baseline for sampling noise alongside a permutation test to construct a bias-corrected null distribution \autocite{ho_ting_bosco_earth_2026}. The observed distance is then evaluated against this null distribution with Benjamini-Hochberg adjusted $p$-values \autocite{benjamini_controlling_1995}. 

Considering the sparsity issue, besides addressing this using ground distance, we also compare the \textit{marginal} distributions of attribute-level choices. For each attribute level $a$, we compute the empirical choice frequency for humans $p_H(a)$ and synthetic agents $p_A(a)$. We then assess the similarity of the distribution using the following metrics. First, we average the Hellinger distances between human and synthetic marginal distributions calculated for each attribute $j \in J$ \autocite{hellinger_neue_1909}: 
\begin{equation}
H_{\text{marg}}(p_H, p_A) = \frac{1}{J} \sum_{j=1}^{J} \frac{1}{\sqrt{2}} \sqrt{\sum_{m=1}^{M_j} \left(\sqrt{p_{H,j}(a_m)} - \sqrt{p_{A,j}(a_m)} \right)^2}
\end{equation}
where $M_j$ is the number of levels for attribute $j$. Second, we compute Pearson's correlation between human and synthetic choice shares across all $M$ attribute levels \autocite{pearson_vii_1895}: \begin{equation} r = \frac{\sum_{m=1}^{M} (p_{H}(a_m) - \bar{p}_H) (p_{A}(a_m) - \bar{p}_A)}{\sqrt{\sum_{m=1}^{M} (p_{H}(a_m) - \bar{p}_H)^2} \sqrt{\sum_{m=1}^{M} (p_{A}(a_m) - \bar{p}_A)^2}} \end{equation} where $\bar{p}_H$ and $\bar{p}_A$ are the respective mean choice shares. Confidence intervals (CIs) are computed for each of these measures via bootstrapping. 

To evaluate the extent to which responses generated by individual synthetic agents conform to expected response distributions, we use weighted F1 scores:
\begin{equation}
\text{Weighted F1} = \sum_{c=1}^C \omega_{c} \times 2\frac{(\text{Precision}_c \times \text{Recall}_c)}{(\text{Precision}_c + \text{Recall}_c)}      
\end{equation}
where $C$ is the number of observed choice classes and $\omega_c$ is the empirical support of class $c$ in human data. This individual-level diagnostic is the strictest among all.

To account for within-cluster correlation, both bootstrapping and permutation are conducted at the respondent level instead of the profile level.  

\subsubsection{Similarity}
\label{test:similarity}
Next, we use AMCEs, MMs, and, when applicable, cAMCEs or other heterogeneous estimands to evaluate the degree of similarity between synthetic and human responses as specified in Hypothesis~\ref{hyp:similarity}.\footnote{Some heterogeneous treatment effect estimations may require information about responses to other questions included in the survey, which require additional simulations beyond demographic profiles that are omitted. Also, the original study may use different estimands. See Appendix~\ref{sec:estimand} for more details.} To do so, we estimate sample coefficients $\hat{\beta}$, which correspond to each of our primary population estimands, and cluster standard errors at the respondent level. For the two studies whose original analyses applied survey weights---\textcite{hainmueller_hidden_2015, bechtel_mass_2013}---we estimate these weighted specifications via weighted least squares using the original weights.

Since conjoint designs include many attribute levels, we utilize the following summary measures for each study. First, we compute Pearson's correlations between the estimate vectors to assess pattern alignment. Second, we compute the root mean squared error (RMSE) of the differences of estimates across all $M$ attribute levels: \begin{equation} \text{RMSE} = \sqrt{ \frac{1}{M} \sum_{m=1}^{M} \left( \hat{\beta}_m^{A} - \hat{\beta}_m^{H} \right)^2 } \end{equation} Third, we compute the proportion of attribute levels where the estimates have the same direction. Fourth, we compute the proportion of synthetic point estimates falling within the corresponding human 95\% CIs as an uncertainty-based diagnostic of statistical correspondence.

\subsubsection{Stability}
\label{test:stability}
Finally, for each distinct analytic choice setting $g \in \{1, 2, ..., G\}$, we estimate the estimands of interest for each attribute level $m$, $\hat{\beta}_{m, g}$ using a variance–decomposition approach that distinguishes between variability due to sampling error and variability attributable to analytic choices. Specifically, we repeat the variance-decomposition analysis separately for model, temperature, and model-by-temperature combinations.

For each attribute level $m$ and each estimand, we fit this random effects model:
\begin{equation}
\hat{\beta}_{m,g} = \alpha_m + u_{m,g} + e_{m,g}
\end{equation}
where $\alpha_m$ is the pooled effect for attribute level $m$ across all settings, $u_{m,g} \sim \mathcal{N}(0, \tau_m^2)$ captures between-setting variability, and $e_{m,g} \sim \mathcal{N}(0, SE_{m,g}^2)$ captures sampling uncertainty. We then compute the stability ratios for each attribute level: \begin{equation} R_m = \frac{\tau_m}{\sqrt{\frac{1}{G} \sum_{g=1}^G SE_{m,g}^2}} \end{equation} where $R_m=1$ provides an intuitive threshold, above which between-setting variability exceeds typical within-setting estimation uncertainty. We generate bootstrap CIs for both $\tau_m^2$ and $R_m$ using $B=2000$.

\subsubsection{Summary}
Table~\ref{tab:test-summary} summarizes the statistical tests for each hypothesis. The thresholds used for testing Hypothesis~\ref{hyp:distribution} follow conventional choices and have been pre-registered. The thresholds for Hypothesis~\ref{hyp:similarity} follow the same logic, except for RMSE, which uses a threshold of 0.05.\footnote{\textcite{schuessler_power_2020} analyzed 15 highly cited forced-choice conjoint experiments and found that the median of their AMCEs is 0.05. An RMSE of 0.05 in this case can be seen as substantively large, as the average discrepancy between synthetic and human estimates is comparable to the typical human treatment effect.}

\begin{table}[htbp]
\centering
\scriptsize
\caption{Summary of Statistical Tests}
\label{tab:test-summary}
\begin{tabularx}{\textwidth}{p{2cm} p{2.3cm} X p{2.3cm} p{4cm}}
\toprule
\textbf{Hypothesis} & \textbf{Test Metric} & \textbf{Methodology} & \textbf{Threshold} & \textbf{Substantive Justification} \\
\midrule
\textbf{Hypothesis~\ref{hyp:distribution}: Distribution} 
& Wasserstein Distance ($W$) 
& Wasserstein Distance using Manhattan distance on one-hot profiles; $B=2000$ permutation test
& $p \ge 0.05$ 
& Observed distance falls within the 95\% CI of human-to-human resampling; indistinguishable from natural sampling noise \\
\addlinespace
& Hellinger Distance ($H$) 
& Distance between marginal choice probability vectors
& 95\% CI upper bound $< 0.3$ 
& Ensures aggregate distributional alignment \\
\addlinespace
& Pearson's Correlation ($r$) 
& Correlation of empirical choice shares across all $M$ attribute levels
& 95\% CI lower bound $> 0.7$ 
& Indicates aggregate preference weight alignment \\
\addlinespace
& Weighted F1-Score 
& Weighted average of profile-level alignment across category classes 
& Score $> 0.7$ 
& Individual-level choices are captured with significantly higher accuracy than random-guess baselines \\
\midrule
\textbf{Hypothesis~\ref{hyp:similarity}: Similarity} 
& Pearson's Correlation ($r$) 
& Correlation between the vectors of human and synthetic AMCEs 
& 95\% CI lower bound $> 0.7$ 
& Assesses strong pattern alignment across the entire set of estimated treatment effects \\
\addlinespace
& Root Mean Squared Error (RMSE) 
& Root mean squared difference between $\hat{\beta}_m^{A}$ and $\hat{\beta}_m^{H}$ across $M$ levels
& RMSE $< 0.05$ 
& Replicated effects do not deviate by more than a common median human AMCE (0.05) \\
\addlinespace
& Directional Agreement 
& Proportion of attribute levels where human and synthetic AMCEs share the same sign
& Proportion $> 0.7$ 
& Ensures qualitative alignment in the direction of treatment effects across attributes \\
\addlinespace
& Coverage Rate 
& Proportion of synthetic AMCEs falling within the 95\% CI of human AMCEs
& Proportion $> 0.7$ 
& Indicates the extent to which synthetic point estimates fall within the corresponding human 95\% CIs \\
\midrule
\textbf{Hypothesis~\ref{hyp:stability}: Stability} 
& Stability Ratio ($R_m$) derived from Between-Setting Variance ($\tau_m^2$) 
& Random-effects variance parameter capturing variability across simulation settings ($g$), standardized as a ratio of between-setting standard deviation ($\tau_m$) to root-mean-square within-setting standard error ($SE$). 
& $R_m \le 1$ evaluated via $B=2000$ bootstrap CIs 
& Quantifies simulation stability across variations in analytic choices by scaling the absolute between-setting instability directly against inherent human sampling noise \\
\bottomrule
\end{tabularx}
\end{table}

\section{Results}
\subsection{Distribution}
\begin{figure}[htbp]
    \centering
    \includegraphics[width=\linewidth]{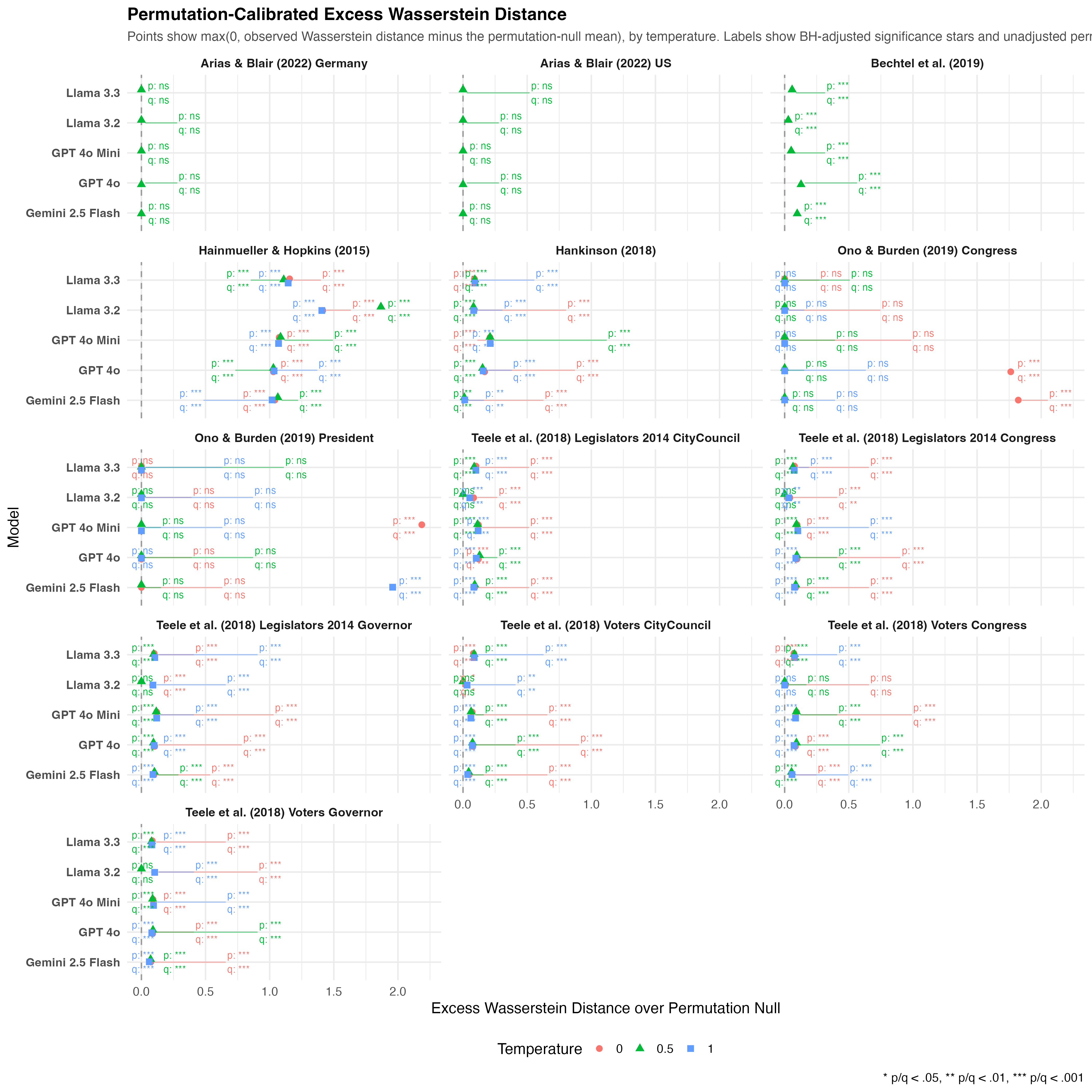}
    \caption{Permutation-Calibrated Excess Wasserstein Distance}
    \label{fig:excess-emd}
\end{figure}

Figure~\ref{fig:excess-emd} reports the permutation-calibrated excess Wasserstein distance.\footnote{Note that non-rejections should not be interpreted as evidence of distributional equivalence. Several of these study designs exhibit high empirical support sparsity, which inflates the finite-sample null distribution and may reduce the power of the permutation test to detect genuine divergence. It provides stronger evidence when it rejects the null than when it fails to reject it, where the latter may reflect either substantive similarity or insufficient power.} At the choice level, for \textcite{arias_changing_2022}, all models generate results where model–human distances are broadly comparable to the human-reference baseline. In contrast, for \textcite{bechtel_mass_2013, hankinson_when_2018, hainmueller_hidden_2015}, all models fail to achieve this.

For \textcite{teele_ties_2018}, the results are rather nuanced. When $T=0.5$, Llama 3.2 always produces a choice distribution comparable to human participants. However, when $T$ is lowered to 0.0 or raised to 1.0, except in the Voters Congress panel, it acts similarly to the other models and fails the test in all panels. This implies that their choice distributions significantly deviate from those of human participants at the aggregate level. \textcite{ono_contingent_2019} also show nuanced results. While all models pass the tests when $T=0.5$, Gemini 2.5 Flash fails at other values of $T$ and GPT-4o and its mini version fail at $T=0.0$ for the Congress and President panels, respectively.

\begin{figure}[htbp]
    \centering
    \includegraphics[width=\linewidth]{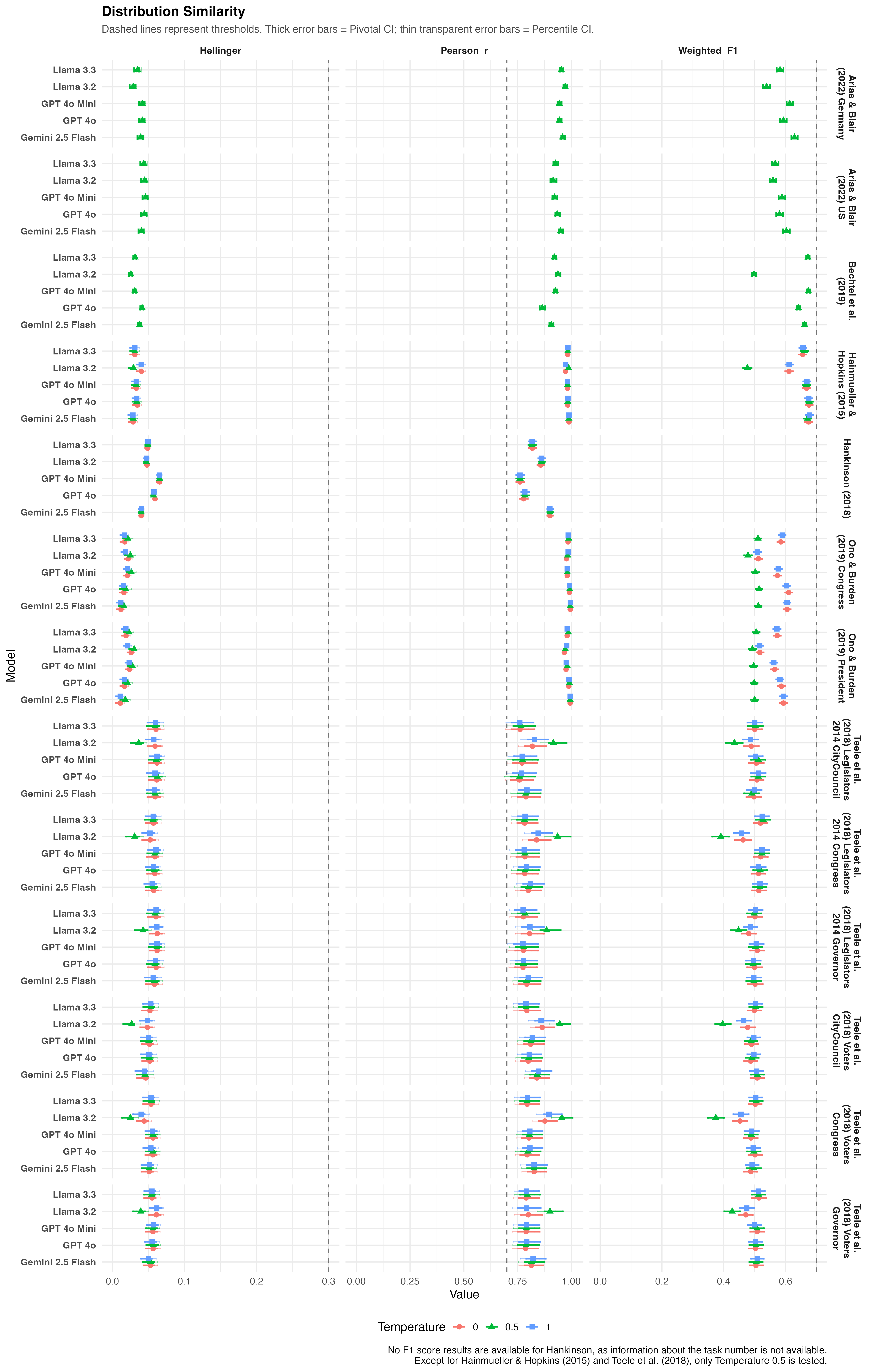}
    \caption{Distribution Similarity}
    \label{fig:h1-metrics}
\end{figure}

Figure~\ref{fig:h1-metrics} reports the results at a lower level. Regarding the similarity of marginal attribute-level choice distributions, across all 13 experimental setups and all 5 models, the $95\%$ bootstrapped CIs for the Hellinger distance and the Pearson $r$ also both satisfy their respective thresholds. At the individual level, none of the combinations pass the test. Unlike the previous cases, Llama 3.2, especially when $T=0.5$, performs worse than the other models.

\subsection{Similarity}
\begin{figure}[htbp]
    \centering
    \includegraphics[width=\linewidth]{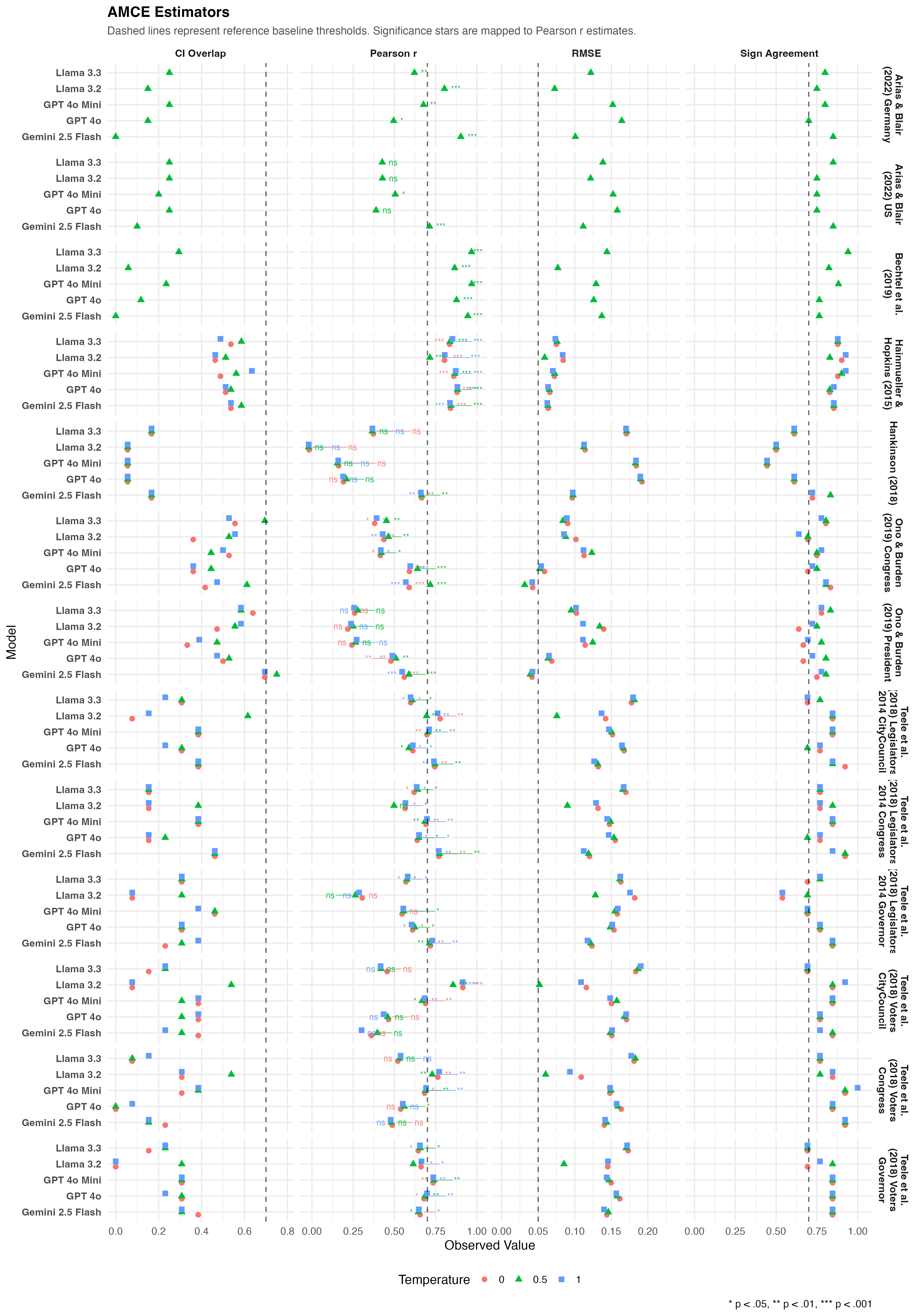}
    \caption{AMCE Estimators}
    \label{fig:h2-amce}
\end{figure}

Figure~\ref{fig:h2-amce} shows the results of the similarity tests for AMCE. For \textcite{hainmueller_hidden_2015, bechtel_mass_2013}, the figures demonstrate strong aggregate-level alignment between human and synthetic causal estimates. They perform exceptionally well in Pearson $r$ and sign agreement. For the other papers, the sign agreement is usually passed or concentrates around the threshold.\footnote{Note that for \textcite{hankinson_when_2018}, the lack of profile order/combination information could possibly affect the results, especially when preferences are relative in conjoint experiments.} The models seem to struggle with Pearson $r$ for cases like \textcite{arias_changing_2022} (US) and \textcite{ono_contingent_2019}, though. There are also multiple instances where the Pearson $r$ result is statistically insignificant, which suggests no correlation between the synthetic and human aggregate preference weights.

Despite occasionally capturing the correct direction and maintaining a high correlation, the proportion of synthetic AMCEs falling within the 95\% CIs of human participants rarely exceeds $0.7$ and the RMSE frequently exceeds $0.05$, which suggests that the models perform less satisfactorily in giving the precise magnitude of human choices, especially for \textcite{bechtel_mass_2013, hankinson_when_2018}. Considering the previous positive finding regarding sign agreement, the case of \textcite{bechtel_mass_2013} provides strong evidence that even if the models get the directions right, the magnitude could be significantly off. However, in the case of \textcite{hainmueller_hidden_2015}, the performance in RMSE across models and temperature settings is much more stable and lower than in the other cases, although none of the cases pass the threshold.  These patterns generally hold true for the case of MM, which allows the derivation of AMCE. See Appendix~\ref{sec:mm} for details.

\begin{figure}[htbp]
    \centering
    \includegraphics[width=\linewidth]{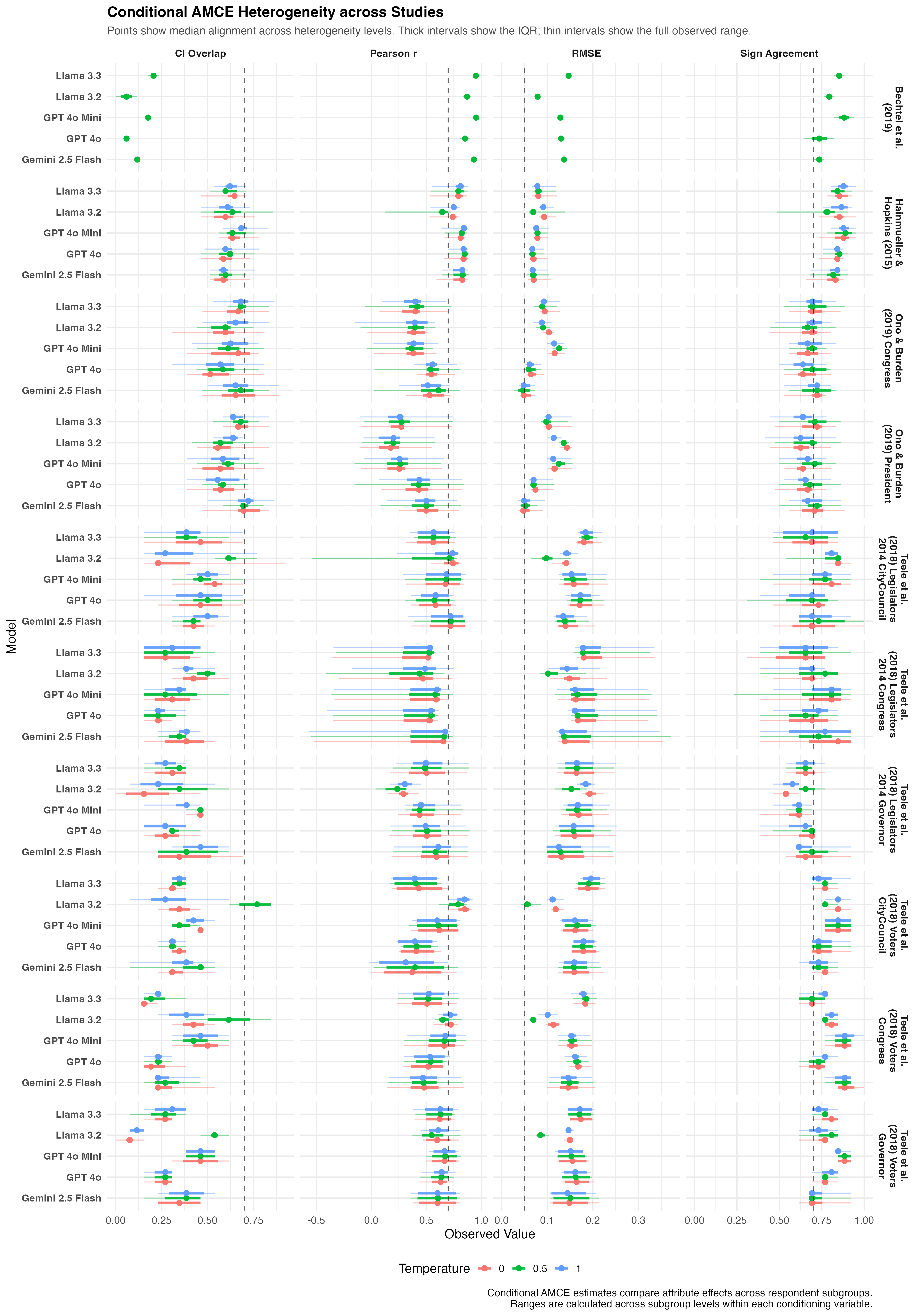}
    \caption{Conditional AMCE Heterogeneity across Studies}
    \label{fig:h2-camce}
\end{figure}

Compared with the aggregate results, the heterogeneity tests are substantially less stable. Figure~\ref{fig:h2-camce} shows that cAMCE alignment is weaker than aggregate AMCE alignment. Although the sign agreement often clusters around the reference threshold, Pearson $r$ and CI overlap are more uneven, while the median RMSE remains relatively high. The weak CI overlap again indicates that the models often approximate the direction of subgroup-specific effects better than their exact magnitude, especially in the case of \textcite{teele_ties_2018} where the sign agreement is strong but the CI overlap is much weaker. These patterns also hold true for the other heterogeneity estimands used by various studies replicated. See Appendix~\ref{sec:heterogeneity} for the discussion.

\subsection{Stability}
While the above shows that the estimates given by synthetic agents are not necessarily similar to the human ones, when we examine the stability of synthetic estimates across analytic settings, the picture appears to be more concerning.

\begin{figure}[htbp]
    \centering
    \includegraphics[width=\linewidth]{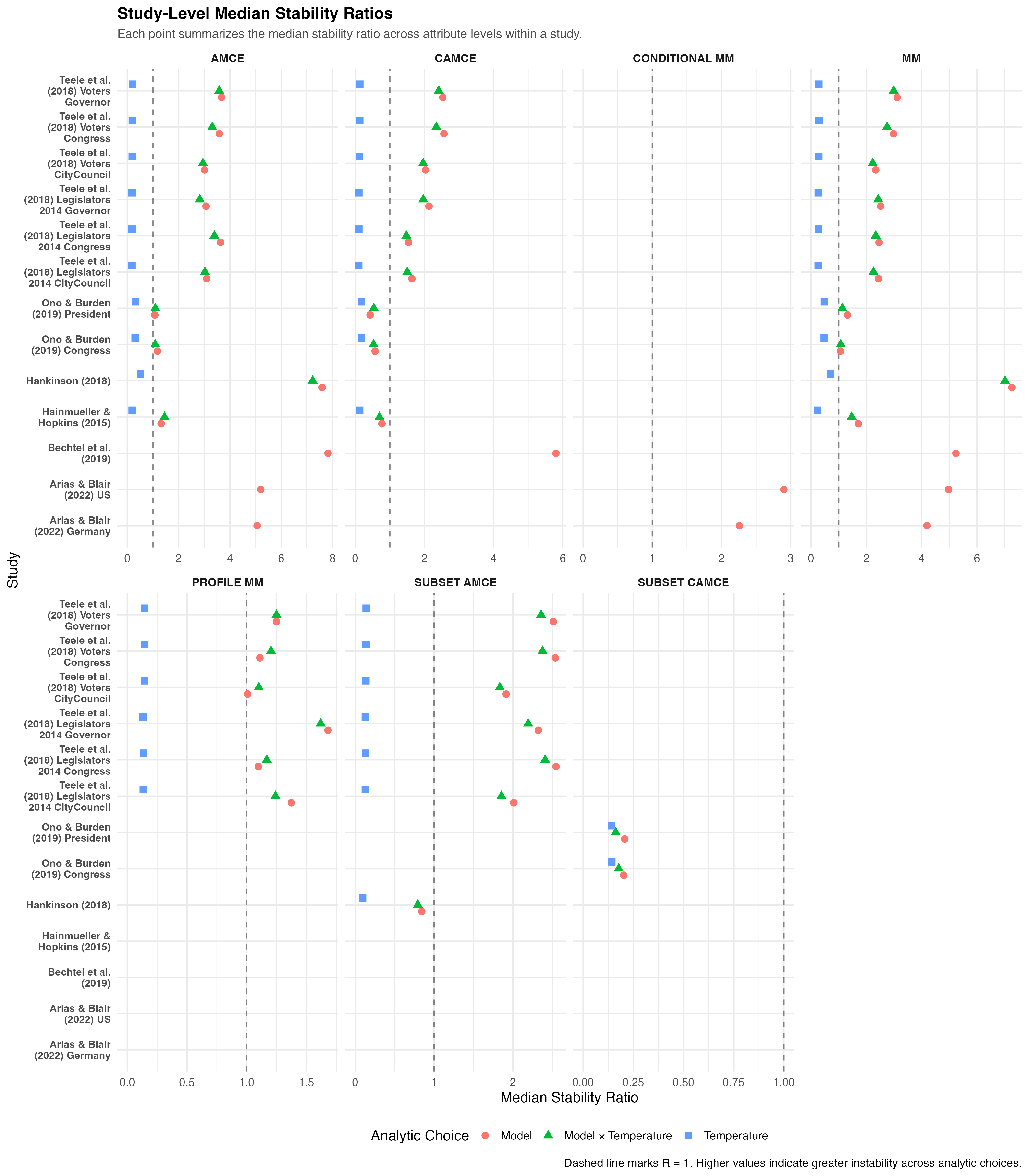}
    \caption{Study-Level Median Stability Ratios}
    \label{fig:median-stability}
\end{figure}

Figure~\ref{fig:median-stability} shows the study-level median stability ratio. Although the degree of instability varies across studies, estimands, and analytic choices, for all studies, many estimates have $R_m>1$ (and sometimes $\gg 1$) when we examine stability across model choices and their interactions with temperature. This indicates that differences across models are often larger than ordinary estimation uncertainty. In contrast, temperature alone generally contributes much less to instability. For instance, for \textcite{teele_ties_2018}, the temperature-based median $R_m$ is close to zero across the estimands. This suggests that changing temperature has relatively weaker independent influence when the model is held fixed. 

\section{Discussion}
Overall, the results do not lend strong support to any of the hypotheses, thus not supporting a general claim that synthetic agents can replace human participants in conjoint experiments. They do reveal a clear hierarchy in which synthetic agents perform best in broad and aggregate benchmarks, and weaken as the test becomes more demanding.

\subsection{Representational Fidelity}
Regarding the distribution metrics that concern Hypothesis~\ref{hyp:distribution}, while LLMs can fairly accurately mimic human choices at the marginal attribute level (passing Hellinger and Pearson $r$), this success does not extend consistently to the full joint distribution of chosen profiles (Wasserstein) or to individual-level choice alignment (Weighted F1).  

This distinction of various layers of distribution matters because conjoint experiments are not only about whether respondents generally prefer one attribute level over another. They are designed to capture trade-offs among multiple attributes. A model that approximates marginal distributions may still fail to reproduce how respondents combine attributes when making choices. A well-documented tendency of LLMs is to collapse heterogeneous perspectives toward a modal response rather than to reproduce the full dispersion of opinion in the population they are prompted to emulate \autocite{dillion_can_2023}. Marginal frequencies could survive this collapse easily because, for example, a model can get the average preference for a high-skill profile right while still missing how skill trades off against country of origin within individual choices. In contrast, the joint distribution and the individual-level match depend on the covariance structure of choices across attributes, which is precisely what a flattening toward the mode erases. Reporting only marginal diagnostics would therefore have produced a substantially more optimistic and misleading verdict.

Moreover, the information processing mechanism of LLMs could be different from the way humans (selectively) process information in conjoint experiments \autocite{druckman_experimental_2022}. Respondents do not necessarily receive and consider all information across multiple dimensions equally, and the same also applies to their synthetic counterparts. The results therefore suggest that synthetic agents are better understood as imperfect aggregate simulators than as one-to-one substitutes for human respondents. They can mimic some population-level patterns, especially as their training data conveys information about the specific socio-political contexts behind the research topic \autocites{wu_large_2023, argyle_out_2023}, but they do not reliably recover the joint and individual-level structure of human choices \autocite{dillion_can_2023}.

\subsection{Inferential Fidelity}
When zooming from the distribution into the causal estimates addressed by Hypothesis~\ref{hyp:similarity}, the same hierarchical interpretive framework remains valid. Synthetic agents sometimes produce high correlations with human AMCEs and MMs, identifying which attributes are generally attractive or unattractive to respondents. However, they are much less successful in reproducing the precise magnitudes of these effects. The case of \textcite{bechtel_mass_2013} serves as an example that even where sign agreement appears strong, the magnitude of the synthetic estimates can remain far from the human benchmark.

In particular, the results of the heterogeneity estimators suggest that a cautious interpretation is critical when examining the reliability of synthetic agents. Estimators like cAMCE require models not only to recover the average direction of human choices, but also to reproduce how those choices vary across respondent subgroups and conditioning attributes. Although median performance sometimes appears acceptable, the full ranges are often wide, and several study-level estimates show substantial instability. This indicates that synthetic agents may approximate broad directional patterns, but they are less reliable for recovering subgroup-specific preference structures, which again highlights that additional calibration might be needed to capture human heterogeneity \autocite{11303355}.

\subsection{Procedural Stability}
The stability analysis further complicates the case for replacement. Although temperature alone contributes little to instability when the model is held fixed, this does not make synthetic estimates robust. The same conjoint design can produce different conclusions depending on which model is used. This echoes \textcite{cummins_threat_2026} that researchers should pay extra attention to their analytic choices. No model is consistently superior across all studies included, estimands, and evaluation metrics, and some are more sensitive to temperature settings compared to the others. 

Researchers hoping to apply synthetic agents robustly may have to justify their use with a cross-model sensitivity analysis \autocite{cummins_threat_2026}. However, this will add extra costs and potentially weaken the main advantage of synthetic agents in data collection, especially as more advanced models may require additional processing or reasoning time and incur higher costs.

\subsection{Claim-Dependent Validation Hierarchy}
Together, the results support a claim-dependent hierarchy for evaluating proposed uses of synthetic participants. Being able to produce a similar preference distribution does not necessarily imply the ability to provide a similar statistical estimate. Importantly, this hierarchy should not be interpreted as a mandate for individual researchers to conduct redundant human benchmark studies alongside every synthetic experiment. Instead, it suggests that the discipline as a whole would require a systematic mapping of this innovation's boundaries across various levels before considering synthetic agents a robust substitute for human samples.

Within these dimensions, the required empirical justification scales directly with the intended level of analysis and the substantive stakes of the research claim. At the least demanding level, synthetic agents may be used for low-stakes aggregate exploration, such as inspecting whether a design produces broadly plausible patterns. However, when researchers proceed to inferential substitution, in which synthetic data replace human observations for estimating quantities, the validation will be more demanding, requiring accurate magnitudes and uncertainty. If researchers are also concerned with respondent-level replacement preserving individual preference variation, the burden will be even higher. Ultimately, this hierarchy also clarifies why the validity of synthetic results cannot be judged by a single favorable statistic or metric.

\section{Conclusion}
As \textcite[1]{aher_using_2023} write, `After all, if simulating human behavior were easy, there would be no need to run human subject experiments as one could simply simulate the outcomes.' While synthetic agents appear to offer a promising alternative to costly human experiments, this paper's analysis across distribution, similarity, and stability has revealed a nuanced picture and confirmed the value of human respondents. Synthetic agents can often approximate broad marginal patterns, but they are much less reliable when asked to reproduce full joint distributions, individual-level choices, precise effect magnitudes, and subgroup heterogeneity. The results are also unstable across settings.

The primary implication is that synthetic agents should not be used uncritically as straightforward or uncalibrated replacements for human participants in conjoint experiments. Their strongest performance appears at the most aggregated level, where they can sometimes recover the general direction of human preferences. However, conjoint analysis often depends on more demanding estimands, especially when subgroup-specific effects are of interest. For these purposes, directional similarity is not enough. If synthetic estimates are systematically too large, too small, or unstable across models, then they may lead researchers to draw misleading substantive conclusions even when they appear to pass simpler benchmarks. More broadly, the apparent performance of synthetic agents depends on what researchers ask them to reproduce: evidence at a lower validation level cannot justify a claim at a higher one.

Although the overall results encourage researchers to stay cautious when using synthetic agents and making substantive claims, this does not render them without value. In particular, synthetic agents can assist robustness-oriented or exploratory analyses. Considering potential flaws in experimental design and execution, synthetic replications could assist researchers in identifying cases where published human results deserve further investigation. Given LLMs' capacity to reflect aggregate preferences, researchers can also leverage synthetic agents to run simulated pilot iterations to test their proposed theory to get a general sense of the aggregate picture before investing financial resources into human deployment. Moreover, although the evaluation criteria used here are fixed, model capability is not. Our results describe where today's models fall on them rather than setting a limit that better models could not one day surpass. Their value may rise with capability improvement, but researchers should still keep in mind that `replacement' could mean different things and the evaluation criterion should vary with the use case.

This paper has some inherent limitations. First, it focuses on the most typical conjoint design and does not cover vignette experiments or new advances like visual conjoints. Future studies can perform similar inquiries in more nuanced experimental settings, and we also encourage scholars to consider other analytic settings beyond model choice and temperature. Second, this paper does not cover advanced chain-of-thought models. Perhaps asking the models to pause and reason before answering would yield a different answer. However, this could add extra costs and weaken one of the key advantages of using synthetic agents to model human preferences. Third, this paper does not examine the synthetic reasoning responses. Although some models may give the same choices, they may not necessarily reach the conclusion with the same reasoning, which could be crucial to theory-testing and examining internal contradictions. Future studies can consider examining the convergence of the open-ended responses using quantitative text analysis \autocite{robertsStructuralTopicModels2014}. 

Finally, a caution concerns the source of the aggregate successes by synthetic agents. The headline findings of the studies replicated are plausibly present in the models' training corpora. Accordingly, the strong aggregate alignment we observe for cases such as \textcite{hainmueller_hidden_2015} may reflect recall of documented results rather than genuine simulation of how the underlying respondents would have chosen. However, this contamination problem ultimately strengthens our conclusions about the limitations of synthetic agents. This mechanism should push synthetic output toward the human benchmark rather than away from it, but synthetic agents are still far from capable of being a robust substitute for human sample. Our reported aggregate performance is then better read as an upper bound than as a typical expectation, thus offering additional evidence that the discipline must pay extra caution when using synthetic agents. Future methodological evaluations may attempt to launch a synthetic test alongside their original conjoint experiment, before the findings enter the public record, to assess whether synthetic agents can genuinely represent human preferences beyond what the training data has recorded.

\section{References}
\printbibliography[heading=none]

\appendix
\counterwithin{figure}{section}
\counterwithin{table}{section}
\counterwithin{equation}{section}

\section{Appendix: Overview of Estimands}
\label{sec:estimand}
Table~\ref{tab:study_estimands} shows an overview of the original estimands and heterogeneity variables of the original studies. Some heterogeneity analyses cannot be replicated, as they depend on extra variables requiring additional simulations.

\begin{table}[htbp]
\centering
\caption{Overview of Original Estimands and Heterogeneity Variables across Included Studies}
\label{tab:study_estimands}
\scriptsize
\begin{tabularx}{\textwidth}{p{2.5cm} p{2.5cm} X X X} 
\toprule
\textbf{Study} & 
\textbf{Original Main Estimand} & 
\textbf{Original Heterogeneity Estimand} & 
\textbf{Included Heterogeneity Variables or Subsets} & 
\textbf{Excluded Heterogeneity Variables or Subsets} \\ 
\midrule

\textcite{arias_changing_2022} & 
AMCE and/or MM for conjoint profiles &
Subgroup marginal means & 
Empathy mean/quartile; age mean/quartile; education college; employed; unemployed; US partisanship & 
US border-state indicators; US urban indicators; Germany East/West state indicators; Germany urban indicators; state-region indicators \\ \addlinespace

\textcite{bechtel_mass_2013} & 
Survey-weighted AMCE and MM &
Interaction-style heterogeneity & 
Left/right ideology & 
Support for international environmental cooperation; reciprocity support / related attitudinal moderators \\ \addlinespace

\textcite{hainmueller_hidden_2015} & 
Post-stratification--weighted AMCE, with restricted-design adjustment for constrained attribute combinations & 
Interaction-based conditional AMCE / AICE & 
Education; party ID; income; White/non-white; Hispanic/not Hispanic; ideology; gender; age & 
Ethnocentrism; industry/fiscal exposure; ZIP diversity; immigration attitude \\ \addlinespace

\textcite{hankinson_when_2018} & 
AMCE & 
AMCE estimated inside filtered subsets & 
Homeowner/renter $\times$ market-rate/affordable; city-rent quintiles $\times$ affordability; ZIP-rent quintiles $\times$ affordability; employed renters $\times$ rent-burden quintiles $\times$ affordability; homeowner income split; homeowner ideology split & 
Price anxiety / city-interest subset; Proposition I ban-support subset; citywide supply-support subset \\ \addlinespace

\textcite{ono_contingent_2019} & 
AMCE & 
Interaction-based conditional AMCE / AICE; subset conditional AMCE & 
Respondent sex; education; age group; class; region; race; partisanship; same-party vs different-party pairings $\times$ respondent partisanship & 
N/A \\ \addlinespace

\textcite{teele_ties_2018} & 
AMCE & 
Respondent-level heterogeneity; candidate-gender subset AMCE; social-role profile marginal means & 
Respondent gender; respondent party; candidate gender subsets; candidate gender $\times$ marriage/spouse $\times$ children social-role profile category & 
Respondent age; respondent education; respondent income \\ 

\bottomrule
\end{tabularx}
\end{table}

In the following, we define the heterogeneous estimands other than cAMCE. First, similar to the case of cAMCE, conditional MM fixes $A_{ijk[C]}=\mathbf{a}_{C}$ and then averages over the remaining profile attributes and the paired profile. It is defined as:
\begin{multline}
    \mu^{(l)}_c(a_1 \mid \mathbf{A}_{ijk[C]} = \mathbf{a}_C) = \sum_{(\mathbf{a}_{R}, \mathbf{a}') \in \tilde{\mathcal{A}}_{R}'} \Big\{\mathbb{E}[Y_{ijk} \mid A_{ijkl} = a_1, \mathbf{A}_{ijk[C]} = \mathbf{a}_{C}, \\
    \mathbf{A}_{ijk[R]} = \mathbf{a}_{R}, \mathbf{A}_{ij[k']} = \mathbf{a}'] \Big\} \cdot \rho(\mathbf{a}_{R}, \mathbf{a}')
\end{multline}
where $R$ denotes the remaining focal-profile attributes not including the focal attribute $l$ and the conditioning attribute set $C$.

Second, profile MM fixes the entire $\mathbf{A}_{ijk}=\mathbf{a}$ instead of fixing one attribute level. It is defined as: 
\begin{equation}
\nu(\mathbf{a}) = \sum_{\mathbf{a}' \in \mathcal{A}} \Big\{\mathbb{E}[Y_{ijk} \mid \mathbf{A}_{ijk} = \mathbf{a}, \mathbf{A}_{ij[k']} = \mathbf{a}'] \Big\} \cdot \rho(\mathbf{a}')
\end{equation}

Third, subset AMCE is defined as the AMCE estimated only within a respondent subset $S$:
\begin{multline}
\pi^{(l)}_{S}(a_1, a_0)= \sum_{(\mathbf{a}_{-l}, \mathbf{a}') \in \tilde{\mathcal{A}}} \Big\{\mathbb{E}[Y_{ijk} \mid A_{ijkl} = a_1, \\
\mathbf{A}_{ijk[-l]} = \mathbf{a}_{-l}, \mathbf{A}_{ij[k']} = \mathbf{a}', i \in S] \\
- \mathbb{E}[Y_{ijk} \mid A_{ijkl} = a_0, \mathbf{A}_{ijk[-l]} = \mathbf{a}_{-l}, \mathbf{A}_{ij[k']} = \mathbf{a}', i \in S]\Big\} \cdot \rho(\mathbf{a}_{-l}, \mathbf{a}')
\end{multline}

Fourth, subset cAMCE is defined as the cAMCE estimated only within a respondent subset $S$:
\begin{multline}
\pi^{(l)}_{c,S} (a_1, a_0 \mid \mathbf{A}_{ijk[C]} = \mathbf{a}_C) \\ = \sum_{(\mathbf{a}_{R}, \mathbf{a}') \in \tilde{\mathcal{A}}_{R}'} \Big\{\mathbb{E}[Y_{ijk} \mid A_{ijkl} = a_1, \mathbf{A}_{ijk[C]} = \mathbf{a}_{C}, \mathbf{A}_{ijk[R]} = \mathbf{a}_{R}, \mathbf{A}_{ij[k']} = \mathbf{a}', i \in S] \\
- \mathbb{E}[Y_{ijk} \mid A_{ijkl} = a_0, \mathbf{A}_{ijk[C]} = \mathbf{a}_{C}, \mathbf{A}_{ijk[R]} = \mathbf{a}_{R}, \mathbf{A}_{ij[k']} = \mathbf{a}', i \in S] \Big\} \cdot \rho(\mathbf{a}_{R}, \mathbf{a}')
\end{multline}

\section{Appendix: Sample Generation}
\subsection{Prompting}
Following typical prompt engineering strategies, we start by defining the context (subject to the availability of the details of the original experimental design and survey questions) and the role of the LLM, providing the instructions for each model to adopt the persona of a given human respondent completing a given conjoint task, explaining the task, and elaborating on the requirements. We mirror the flow of a conjoint experiment embedded in an online survey where participants are presented with repeated choice tasks one at a time.

An example of the full structure of the system prompt for setting the context is shown below:

\texttt{This is a technical simulation exercise of a formal academic survey experiment and a valid response ALWAYS exists and MUST be generated. In a discrete choice experiment, respondents have to choose between hypothetical, multi-attribute options. Based on stated preferences, it forces trade-offs to determine how people value specific attributes. The idea is to understand the effects of the attributes by iterating multiple rounds of tasks. \\ \\ In this exercise, you are simulating a data point of a pre-defined respondent participating in a specific round of tasks in a discrete choice experiment to understand preferences towards a political science topic for academic research purposes ONLY. Despite the role-play language, you are ONLY generating a synthetic survey response record for a hypothetical respondent in an academic research dataset for technical simulation. You are NOT making a real decision and you are NOT providing advice. The goal is NOT to express an AI opinion, but to simulate human decision-making analytically. Refusal to choose prevents the completion of the scientific simulation. \\ \\ The participant you will simulate has the following demographic profile. Use only knowledge that they would have. \\ --- DEMOGRAPHICS START \\ You are a 35 year old man living in California, with college education. You have a self-reported middle income and identify yourself as having a centrist political identity. \\ --- DEMOGRAPHICS END ---  \\ \\ --- STUDY CONTEXT START --- \\ This study considers immigration and who is permitted to come to the United States to live. For the next few minutes, we are going to ask you to act as if you were an immigration official. We will provide you with several pieces of information about people who might apply to move to the United States. For each pair of people, please indicate which of the two immigrants you would personally prefer to see admitted to the United States. This exercise is purely hypothetical. Please remember that the United States receives many more applications for admission than it can accept. Even if you aren’t entirely sure, please indicate which of the two you prefer.\\ --- STUDY CONTEXT END ---\\ \\ --- EXAMPLE START --- \\ {[}Task Introduction] \\ Please read the descriptions of the potential immigrants carefully. Then, please indicate which of the two immigrants you would personally prefer to see admitted to the United States. \\ \\ {[}Input Data] \\ === PROFILE 1 === \\- Prior Trips to the U.S.: Entered the U.S. on a tourist visa \\- Reason for Application: Seek a better job \\- Country of Origin: Canada \\- Language Skills: Spoke fluent English \\- Profession: Teacher \\- Job Experience: Five years of experience \\- Employment Plans: Has a contract with a U.S. employer \\- Education Level: Equivalent to completing a college degree in the U.S. \\- Gender: Female \ \\ \\=== PROFILE 2 ===\\ - Prior Trips to the U.S.: Never entered the U.S. \\- Reason for Application: Reunite with family members already in U.S. \\- Country of Origin: Poland \\- Language Skills: Spoke broken English \\- Profession: Construction worker \\- Job Experience: No experience \\- Employment Plans: Will look for work in the U.S. \\- Education Level: No formal education \\ - Gender: Male \\ \\ {[}Question] \\ If you had to choose between them, which of these two immigrants should be given priority to come to the United States to live?  \\ On a scale from 1 to 7, how much should the United States admit Immigrant 1? (1 = absolutely not admit, 7 = definitely admit) \\ On a scale from 1 to 7, how much should the United States admit Immigrant 2? (1 = absolutely not admit, 7 = definitely admit) \\ --- EXAMPLE END --- \\ \\ Follow the EXACT output format below STRICTLY and DO NOT return anything else. \\ --- OUTPUT FORMAT START --- \\ Choice: {[}Choice Number]\\Rating 1: {[}Rating 1 Value] \\Rating 2: {[}Rating 2 Value] \\ Reasoning: {[}Reasoning] \\ --- OUTPUT FORMAT END --- \\ \\ Return ONLY the output. DO NOT refuse.}

An example of the full structure of the user prompt is shown below:

\texttt{Please read the descriptions of the potential immigrants carefully. Then, please indicate which of the two immigrants you would personally prefer to see admitted to the United States. \\ \\ === PROFILE 1 === \\- Prior Trips to the U.S.: Never entered the U.S. \\- Reason for Application: Reunite with family members already in U.S. \\- Country of Origin: Mexico \\- Language Skills: Spoke basic English \\- Profession: Restaurant worker \\- Job Experience: One to two years of experience \\- Employment Plans: Has interviewed with U.S. employers \\- Education Level: Equivalent to completing two years of college in the U.S. \\- Gender: Female \\ \\ === PROFILE 2 === \\- Prior Trips to the U.S.: Entered the U.S. multiple times on a tourist visa \\- Reason for Application: Reunite with family members already in U.S. \\- Country of Origin: China \\- Language Skills: Spoke fluent English \\- Profession: Restaurant worker \\- Job Experience: More than five years of experience \\- Employment Plans: Has interviewed with U.S. employers \\- Education Level: Equivalent to completing two years of college in the U.S. \\- Gender: Male \\ \\ If you had to choose between them, which of these two immigrants should be given priority to come to the United States to live?  \\ On a scale from 1 to 7, how much should the United States admit Immigrant 1? (1 = absolutely not admit, 7 = definitely admit) \\ On a scale from 1 to 7, how much should the United States admit Immigrant 2? (1 = absolutely not admit, 7 = definitely admit)}

\section{Appendix: Choice Extraction}
Table~\ref{tab:extraction_proportions} shows the proportion of valid choice extraction across studies and models, excluding the ones with perfect valid responses, which account for the majority of cases. The invalid responses comprise reluctance to give a response and some edge cases that are difficult to parse, which only form a minority and we decided to ignore them. In particular, for Llama 3.3, it struggles more to give a valid response and occasionally, it gives only `Choice: PROFILE' and does not indicate its choice.

\begin{table}[htbp]
\centering
\caption{Valid Choice Extraction Proportions across Studies and Models (<100\%)}
\label{tab:extraction_proportions}
\scriptsize
\begin{tabularx}{\textwidth}{l X c c}
\toprule
\textbf{Study} & \textbf{Model} & \textbf{Temperature} & \textbf{Proportion} \\ \midrule
\textcite{arias_changing_2022} (Germany) & Llama 3.2 & 0.5 & 0.9942 \\
\textcite{arias_changing_2022} (Germany) & GPT-4o & 0.5 & 0.9999 \\
\textcite{arias_changing_2022} (Germany) & GPT-4o mini & 0.5 & 0.9996 \\
\textcite{arias_changing_2022} (US) & GPT-4o & 0.5 & 0.9999 \\
\textcite{arias_changing_2022} (US) & GPT-4o mini & 0.5 & 0.9994 \\
\textcite{hainmueller_hidden_2015} & Llama 3.2 & 0.5 & 0.9989 \\
\textcite{hainmueller_hidden_2015} & GPT-4o & 1.0 & 0.9981 \\
\textcite{hankinson_when_2018} & GPT-4o & 1.0 & 0.9995 \\
\textcite{hankinson_when_2018} & GPT-4o mini & 0.0 & 0.9997 \\
\textcite{ono_contingent_2019} (Congress) & Gemini 2.5 Flash & 1.0 & 0.9994 \\
\textcite{ono_contingent_2019} (Congress) & Llama 3.2 & 0.0 & 0.9907 \\
\textcite{ono_contingent_2019} (Congress) & Llama 3.2 & 0.5 & 0.9816 \\
\textcite{ono_contingent_2019} (Congress) & Llama 3.2 & 1.0 & 0.9716 \\
\textcite{ono_contingent_2019} (Congress) & Llama 3.3 & 0.0 & 0.9929 \\
\textcite{ono_contingent_2019} (Congress) & Llama 3.3 & 0.5 & 0.9832 \\
\textcite{ono_contingent_2019} (Congress) & Llama 3.3 & 1.0 & 0.9924 \\
\textcite{ono_contingent_2019} (Congress) & GPT-4o & 1.0 & 0.9999 \\
\textcite{ono_contingent_2019} (President) & Gemini 2.5 Flash & 1.0 & 0.9997 \\
\textcite{ono_contingent_2019} (President) & Llama 3.2 & 0.0 & 0.9999 \\
\textcite{ono_contingent_2019} (President) & Llama 3.2 & 0.5 & 0.9927 \\
\textcite{ono_contingent_2019} (President) & Llama 3.2 & 1.0 & 0.9828 \\
\textcite{ono_contingent_2019} (President) & Llama 3.3 & 0.0 & 0.9912 \\
\textcite{ono_contingent_2019} (President) & Llama 3.3 & 0.5 & 0.9848 \\
\textcite{ono_contingent_2019} (President) & Llama 3.3 & 1.0 & 0.9909 \\
\textcite{ono_contingent_2019} (President) & GPT-4o & 0.5 & 0.9999 \\
\textcite{ono_contingent_2019} (President) & GPT-4o & 1.0 & 0.9999 \\
\textcite{teele_ties_2018} (Legislators 2014 City Council) & Llama 3.3 & 0.5 & 0.9747 \\
\textcite{teele_ties_2018} (Legislators 2014 City Council) & GPT-4o & 1.0 & 0.9994 \\
\textcite{teele_ties_2018} (Legislators 2014 Congress) & Llama 3.3 & 0.5 & 0.9767 \\
\textcite{teele_ties_2018} (Legislators 2014 Governor) & Llama 3.3 & 0.5 & 0.9805 \\
\textcite{teele_ties_2018} (Voters City Council) & Llama 3.3 & 0.5 & 0.9714 \\
\textcite{teele_ties_2018} (Voters Congress) & Llama 3.3 & 0.5 & 0.9718 \\
\textcite{teele_ties_2018} (Voters Governor) & Llama 3.3 & 0.5 & 0.9748 \\ \bottomrule
\end{tabularx}
\end{table}

\section{Appendix: Wasserstein Distance Calibration}
We use a permutation-based null calibration to account for the inherent upward bias of the Wasserstein distance and avoid over-estimating the distribution difference, especially in the case of sparse data structures prevalent in conjoint experiments. Table~\ref{tab:Wasserstein_comparison_pattern} reports the differences in the results. While a more conservative method is used, most cases still show significant distribution differences as the main body shows.

\begin{table}[htbp]
\centering
\caption{Comparison of Divergence Flags Between Raw and Permutation-Calibrated Wasserstein Distance}
\label{tab:Wasserstein_comparison_pattern}
\begin{tabularx}{\textwidth}{X r r}
\toprule
\textbf{Comparison Pattern} & \textbf{\textit{n}} & \textbf{Percent} \\
\midrule
Both flag model-human divergence & 121 & 73.3\% \\
Neither flags divergence & 39 & 23.6\% \\
Raw suggests divergence; permutation does not & 5 & 3.03\% \\
\bottomrule
\end{tabularx}
\end{table}

We acknowledge that under sparse empirical support, the power of the calibration remains limited. This implies that true population divergence could be virtually indistinguishable from baseline sampling noise. To evaluate whether our Wasserstein results are driven by support sparsity, we follow the diagnostic recommendations by \textcite{ho_ting_bosco_earth_2026} to analyze the observation-to-observed-support ratio, $N/|\hat{\mathcal{S}}|$, together with the frequency distribution of observed support points across all 165 replication runs. In addition to the proportion of support points occurring five times or fewer, we report the proportion occurring exactly once (singletons) and no more than twice. These additional thresholds allow us to assess whether the diagnosis of sparsity depends on the particular choice of five occurrences as the low-frequency cutoff.

As shown in Table~\ref{tab:support_diagnostics}, 12 out of the 13 replicated study setups exhibit effective empirical support sparsity. Across these setups, the mean observation-to-support ratio ranges from 1.00-2.93, while between 90.49\%-100\% of observed support points occur five times or fewer.\footnote{These diagnostics characterize the observed empirical support, $\hat{\mathcal S}$, rather than the complete theoretical or combinatorial support of each experimental design, $K_\text{theoretical}$. Some high-dimensional designs can have $K_\text{theoretical} \gg N$ even when $N/|\hat{\mathcal S}|$ is close to one. For example, in \textcite{ono_contingent_2019}, almost every realized profile is unique, and thus gives a $N/|\hat{\mathcal S}| \approx 1$. This reflects an extremely diffuse empirical distribution over a much larger underlying profile space rather than dense coverage of that space. Thus, $N/|\hat{\mathcal S}|$ measures the average number of observations per empirically observed support point and should not be interpreted as the proportion of the theoretically possible profile space covered by the sample.} Sparsity remains rather substantial under considerably stricter definitions of low frequency. In \textcite{arias_changing_2022}, for example, approximately 48\% of observed support points are singletons and approximately 77\% occur no more than twice. Across the \textcite{teele_ties_2018} and \textcite{hankinson_when_2018} setups, approximately 25-32\% of support points are singletons and 49-58\% occur no more than twice, despite their somewhat higher $N/|\hat{\mathcal{S}}|$ ratios of 2.60-2.93. The only exception is \textcite{bechtel_mass_2013}, which features $34,000$ observations spread over $2,812$ unique profiles ($N / |\mathcal{S}| = 12.09$). Only 2.94\% of its observed support points are singletons, 6.29\% occur no more than twice, and 18.71\% occur five times or fewer. 

\begin{table}[htbp]
\centering
\caption{Empirical Support Diagnostics across Replicated Studies}
\label{tab:support_diagnostics}
\scriptsize
\begin{tabularx}{\textwidth}{X r r r r r r}
\toprule
\textbf{Study} &
\textbf{Mean $N$} &
\textbf{Mean $|\hat{\mathcal S}|$} &
\textbf{$N/|\hat{\mathcal S}|$} &
\textbf{Singleton \%} &
\textbf{$\leq2$ \%} &
\textbf{$\leq5$ \%} \\
\midrule
\textcite{arias_changing_2022} (Germany)
& 10,417.8 & 5,556.4 & 1.88 & 47.68 & 77.07 & 99.32 \\
\textcite{arias_changing_2022} (US)
& 10,762.6 & 5,746.2 & 1.87 & 48.07 & 77.54 & 99.07 \\
\textcite{bechtel_mass_2013}
& 34,000.0 & 2,812.2 & 12.09 & 2.94 & 6.29 & 18.71 \\
\textcite{hainmueller_hidden_2015}
& 7,954.4 & 7,003.9 & 1.14 & 99.55 & 99.99 & 99.99 \\
\textcite{hankinson_when_2018}
& 15,094.2 & 5,155.7 & 2.93 & 24.82 & 48.81 & 90.86 \\
\textcite{ono_contingent_2019} (Congress)
& 7,868.3 & 7,866.7 & 1.00 & 99.98 & 100.00 & 100.00 \\
\textcite{ono_contingent_2019} (President)
& 7,884.3 & 7,880.9 & 1.00 & 99.96 & 100.00 & 100.00 \\
\textcite{teele_ties_2018} (Legislators 2014 City Council)
& 1,776.9 & 683.9 & 2.60 & 31.67 & 58.11 & 93.70 \\
\textcite{teele_ties_2018} (Legislators 2014 Congress)
& 1,841.1 & 691.7 & 2.66 & 29.85 & 56.57 & 92.84 \\
\textcite{teele_ties_2018} (Legislators 2014 Governor)
& 1,791.7 & 682.6 & 2.63 & 31.43 & 56.70 & 93.41 \\
\textcite{teele_ties_2018} (Voters City Council)
& 2,060.1 & 726.1 & 2.84 & 25.99 & 51.47 & 91.07 \\
\textcite{teele_ties_2018} (Voters Congress)
& 2,052.1 & 725.9 & 2.83 & 26.41 & 51.27 & 91.34 \\
\textcite{teele_ties_2018} (Voters Governor)
& 2,101.5 & 723.1 & 2.91 & 25.56 & 50.07 & 90.49 \\
\bottomrule
\end{tabularx}
\end{table}

Despite this effective support sparsity in 12 of the 13 study setups, the permutation calibration test flags significant model-human joint distributional divergence ($p < 0.05$) in 73.3\% of cases ($n=121$), while only 3.03\% ($n=5$) of cases are flagged as divergent by raw $W_1$ but cleared by permutation. In particular, in \textcite{bechtel_mass_2013} where support density is high and statistical power is relatively higher, synthetic agents fail the Wasserstein test across all models ($W_{\text{excess}} > 0, p < 0.001$). Taken together, these results indicate that the observed model-human divergence is not simply an artifact of sparse empirical support. Significant divergence frequently remains detectable after accounting for the elevated finite-sample baseline, and it persists in the benchmark for which support density and statistical power is substantially greater.

\section{Appendix: Deviation from the Pre-Analysis Plan}
This study was pre-registered at OSF. In the following, we explain some deviations driven by technical or practical concerns and redundancy considerations.

\subsection{Studies Included}
While we only replicate six papers in total, \textcite{clayton_correcting_2026} actually replicate eight studies, including \textcite{blackman_religion_2018} and \textcite{mummolo_why_2017}. However, the replication files of \textcite{blackman_religion_2018} are not available online. As our analysis strategy requires the use of the original data, we unfortunately have to skip it. Similarly, the raw data files of \textcite{mummolo_why_2017} are not available. 

We have attempted to contact the authors of these two papers and also \textcite{hankinson_when_2018} to request additional information about raw data, but we did not receive any responses.

\subsection{Distribution Analysis}
In the PAP, we specified the use of the global Hellinger distance: \begin{equation} H(p_H, p_A) = \frac{1}{\sqrt{2}} \sqrt{ \sum_{a} \left( \sqrt{p_H(a)} - \sqrt{p_A(a)} \right)^2 } \end{equation} which could force the metric toward its upper bound given the sparse joint distribution that the metric was originally intended to address. We corrected this technical issue by switching to the dense marginal distributions of attribute-level choices instead.

Also, while the PAP did not explicitly specify the level of clustering for statistical inference, performing bootstrapping and permutation at the individual choice or profile level violates the independence assumptions required for valid non-parametric inference. To account for this within-cluster correlation and avoid artificially deflated standard errors, we conduct all bootstrapping and permutation procedures at the respondent level rather than the profile level.

\subsection{Estimand}
We originally planned to estimate only AMCE, MM, and cAMCE for all studies, given that these are the main estimands for conjoint studies. However, not all studies included them as the main estimand. Therefore, we decided to estimate AMCE and MM for all studies, and also perform the heterogeneity analysis done by the original papers that is not necessarily built on cAMCE, so as to improve alignment with the original studies' actual designs.

\subsection{Model Choice}
We originally planned to test Gemini 2.5 Pro. However, considering the disadvantages associated with advanced chain-of-thought models discussed in the conclusion, especially when the level of reasoning efforts is set to high, we decided to drop this particular model.

\subsection{Temperature}
We originally planned to test the case of $T=1.5$ following \textcite{cummins_threat_2026}. However, when we tested GPT-4o on \textcite{teele_ties_2018}, the estimated processing time increased by over 10 times, partly due to some random outputs that are hundreds of lines long, which would significantly increase the cost. Occasionally, the reasoning explanation becomes chaotic and includes non-English characters, while the logic does not flow coherently. Therefore, we decided to drop $T=1.5$.

\subsection{Other Analytic Choices}
We originally planned to consider stability concerning the sample size and task length. Regarding the number of participants, we initially planned to generate a master population of $I = 10,000$ participants constructed by multiplying the number of participants in a certain demographic profile by the same constant $\{n \in \mathbb{Z} \mid n \ge 2\}$ until the total number of participants in the pool exceeds 5000, so as to maintain the representative sampling logic. For analyses at lower participant thresholds, we planned to perform representative random subsampling without replacement from this master pool. Regarding the number of tasks, we initially planned to assign each participant a full sequence of $J = 30$ tasks (the upper bound used by \textcite{bansak_number_2018} when investigating survey satisficing) and analyze these tasks using expanding sequential windows in increments of five (e.g., $J \in \{5, 10, \dots, 30\}$). 

However, we decided to omit these analyses due to the following reasons. First, scaling the pool size does not introduce genuine unobserved variance. Instead, it forces the models through repetitive generative pathways across identical prompt inputs. A potentially meaningful scaling pathway would be to pre-define a joint distribution and sample from it when scaling, but this requires making assumptions on the population that the original survey is drawing from. Second, synthetic agents are immune to survey satisficing, so testing expanding sequential windows fails to mirror real-world survey dynamics. It only measures the context window mechanics and token-tracking stability of the underlying model. Third, scaling the sample would introduce prohibitive computational overhead and data generation costs, which we did not expect at first. When doing some testing on \textcite{hainmueller_hidden_2015}, we identified issues with the internal safeguard of Llama 3.2, which mistakenly treats the survey prompt as promoting discrimination. We have accordingly come up with a more detailed system prompt and use the same structure for every experiment for consistency, which would, however, significantly inflate the cost. This also led us to only test $T=0.0$ and $T=1.0$ for four of the six studies replicated.

\section{Appendix: Marginal Means Similarity Test}
\label{sec:mm}
\begin{figure}[htbp]
    \centering
    \includegraphics[width=\linewidth]{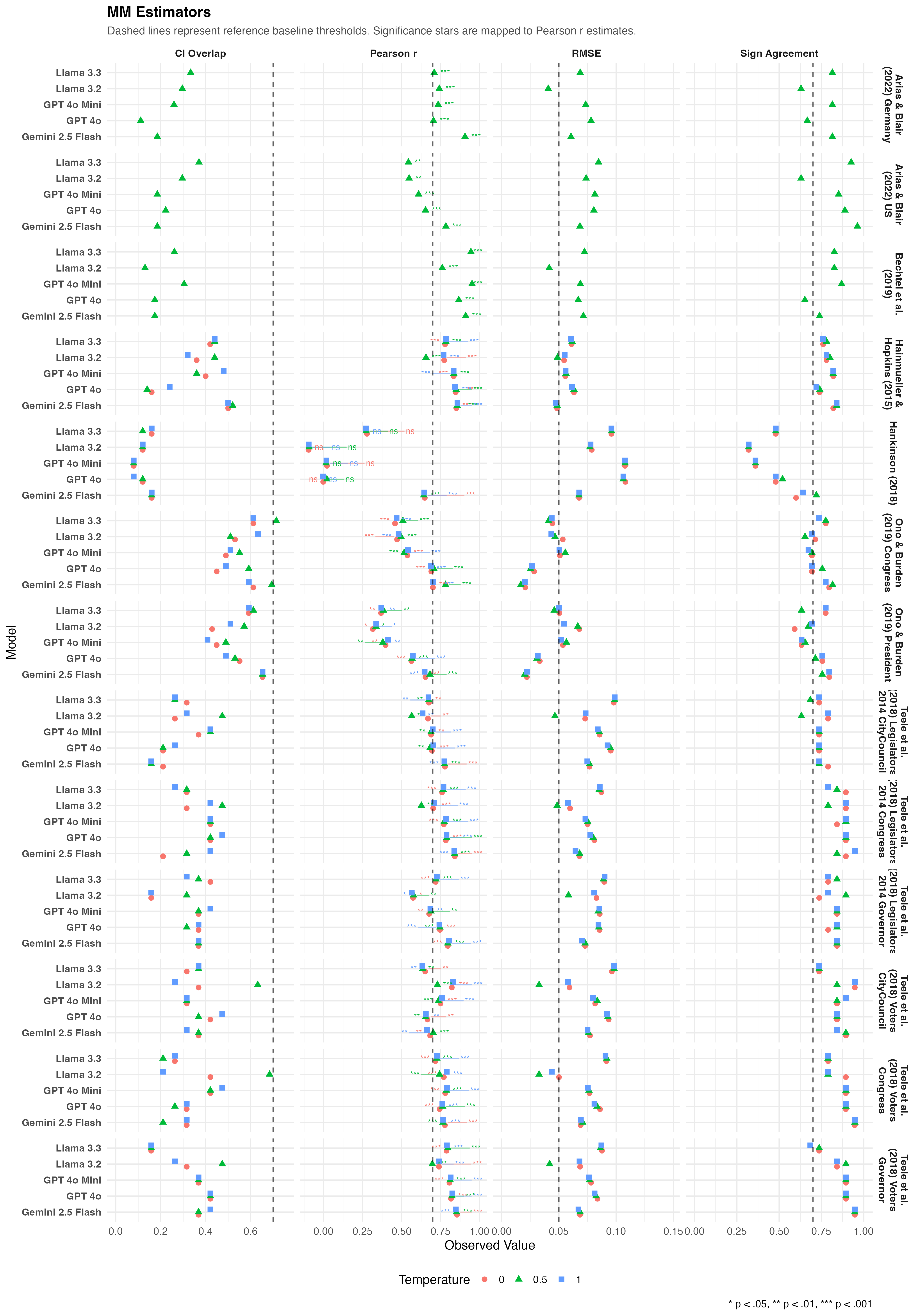}
    \caption{MM Estimators}
    \label{fig:h2-mm}
\end{figure}

Figure~\ref{fig:h2-mm} reports the results of the similarity tests for MMs.

\section{Appendix: Additional Heterogeneity Analysis}
\label{sec:heterogeneity}
\begin{figure}[htbp]
    \centering
    \includegraphics[width=\linewidth]{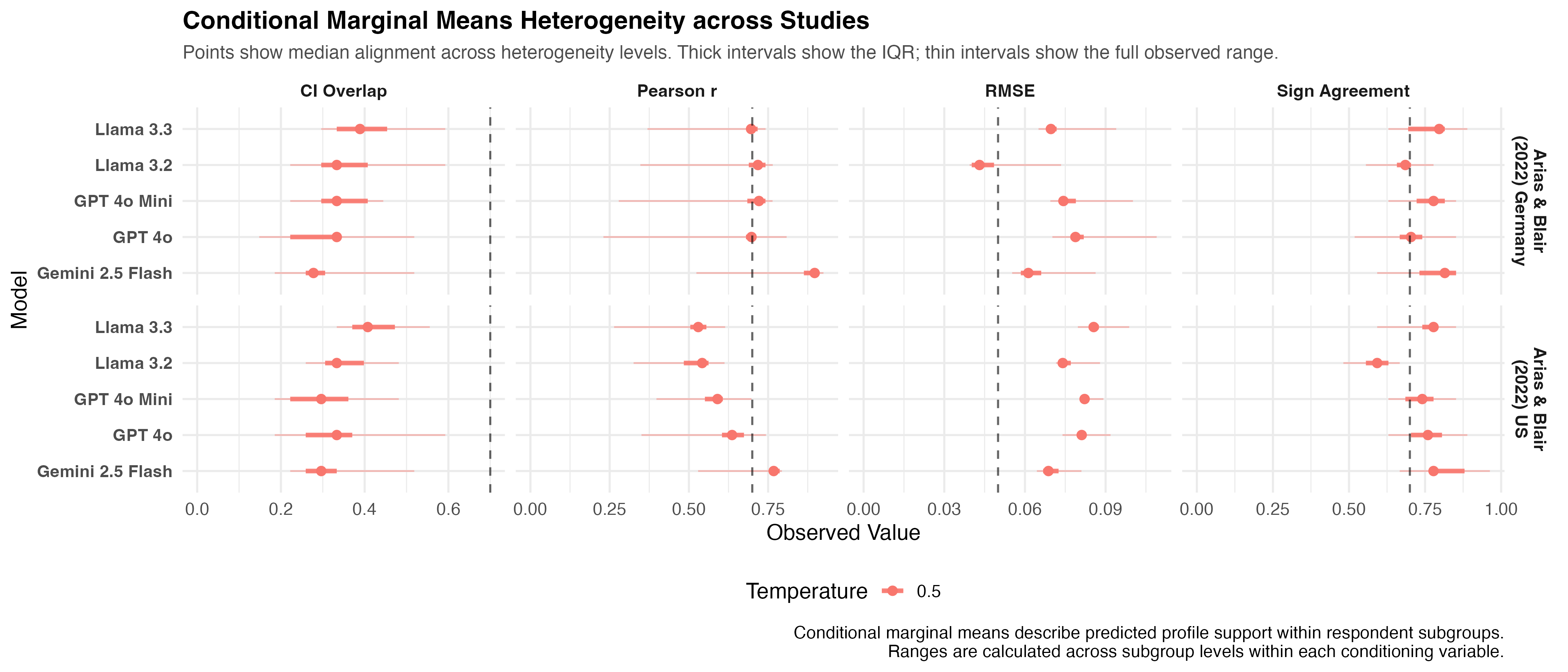}
    \caption{Conditional MM Heterogeneity across Studies}
    \label{fig:h2-cmm}
\end{figure}

\begin{figure}[htbp]
    \centering
    \includegraphics[width=\linewidth]{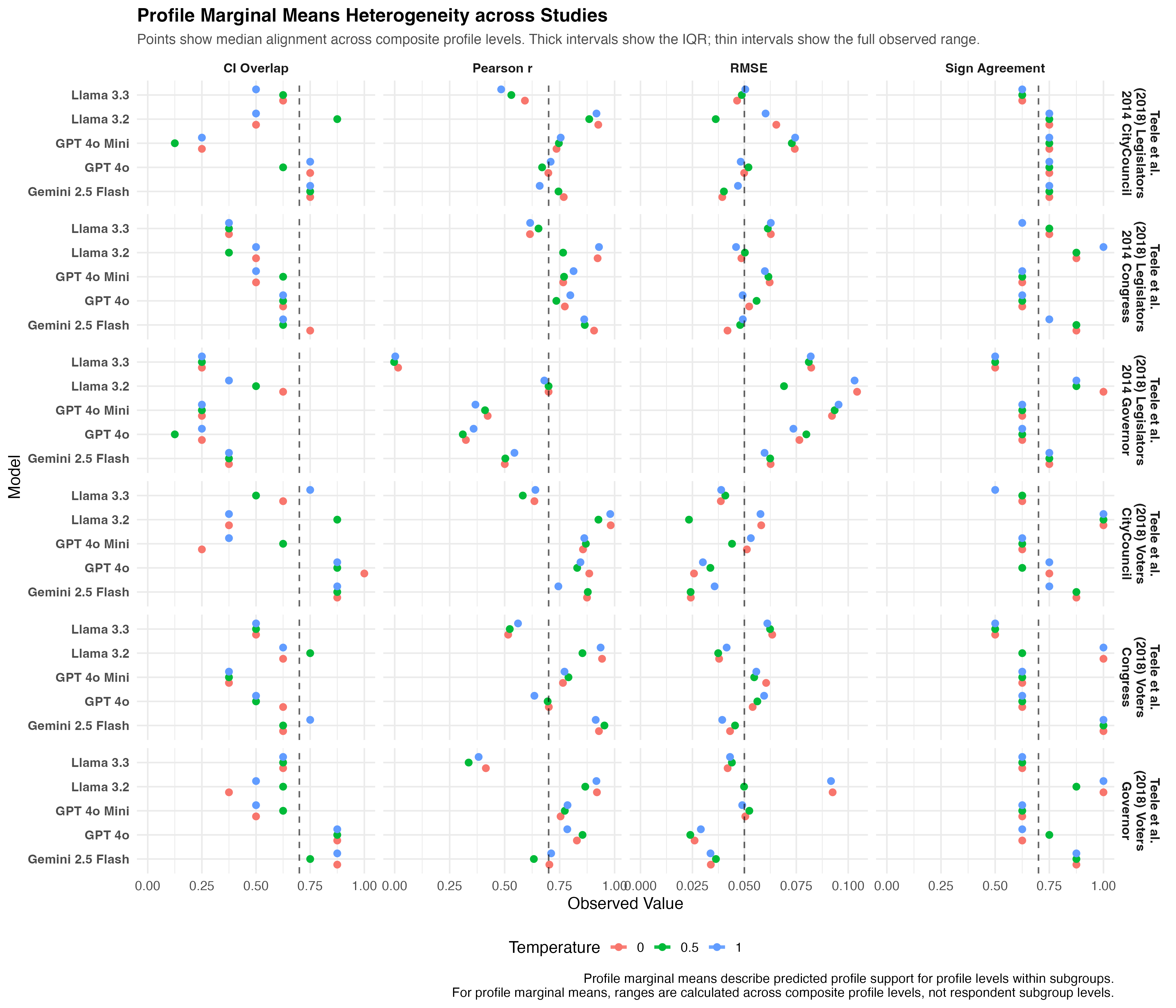}
    \caption{Profile MM Heterogeneity across Studies}
    \label{fig:h2-pmm}
\end{figure}

\begin{figure}[htbp]
    \centering
    \includegraphics[width=\linewidth]{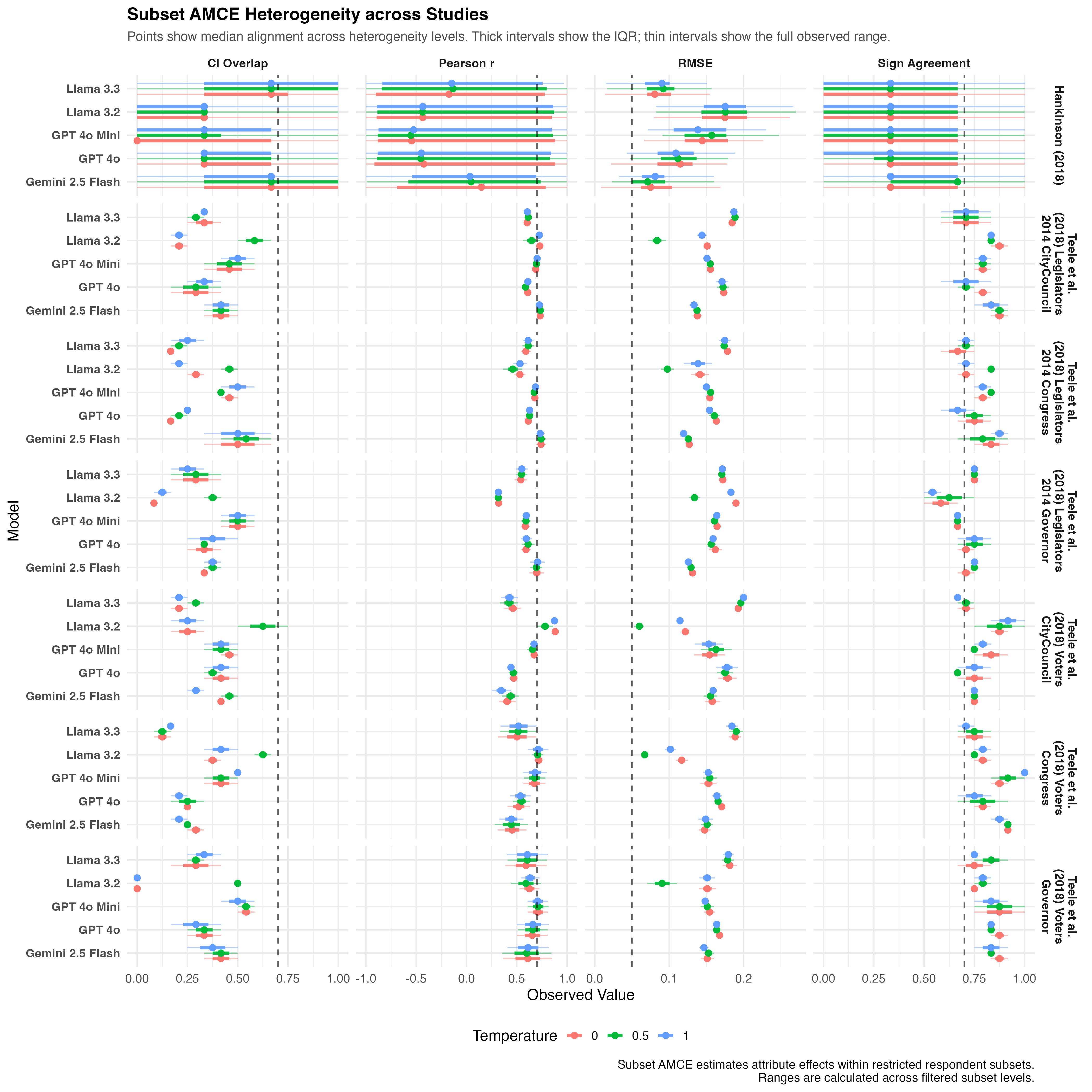}
    \caption{Subset AMCE Heterogeneity across Studies}
    \label{fig:h2-samce}
\end{figure}

\begin{figure}[htbp]
    \centering
    \includegraphics[width=\linewidth]{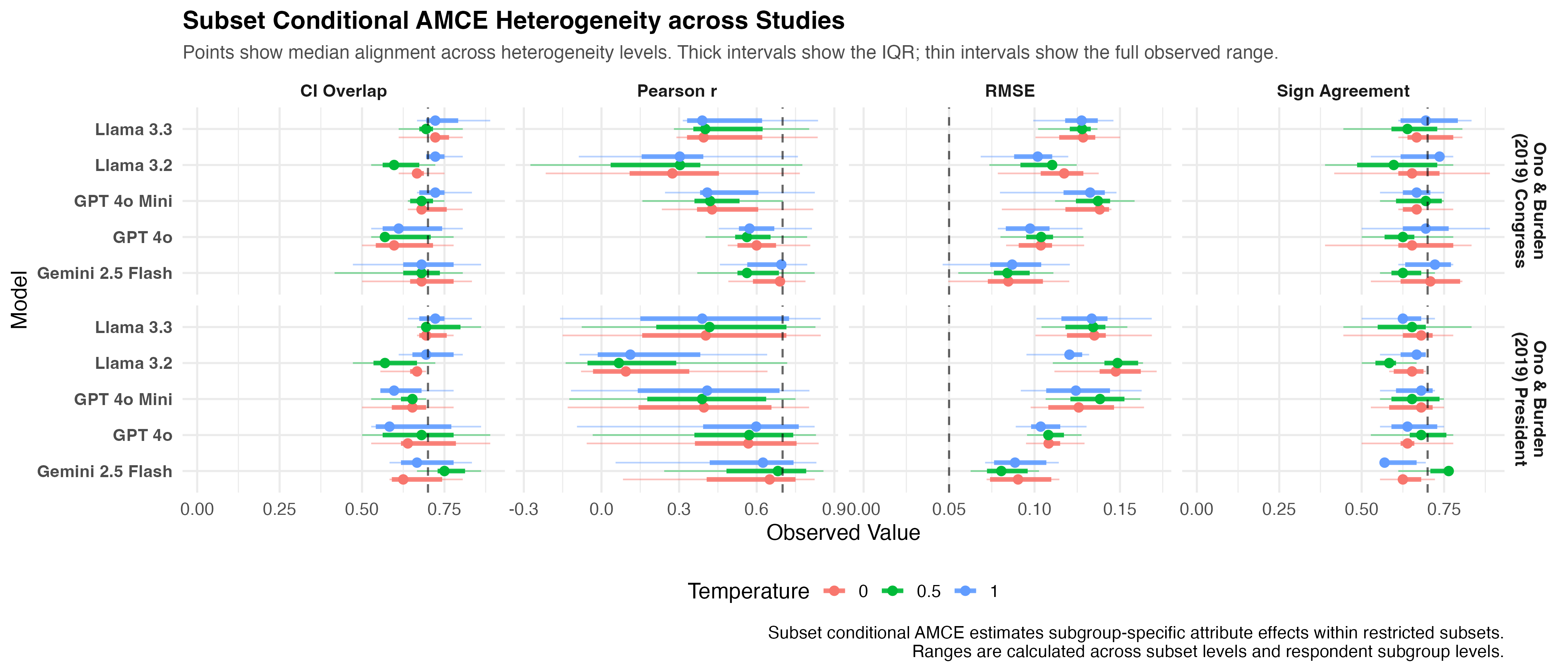}
    \caption{Subset cAMCE Heterogeneity across Studies}
    \label{fig:h2-scamce}
\end{figure}

Figures~\ref{fig:h2-cmm} through \ref{fig:h2-scamce} present the performance of the models across different subgroup heterogeneity specifications beyond cAMCE, including conditional MM, profile MM, subset AMCE, and subset cAMCE. 

A similar phenomenon of relatively strong aggregate alignment and weak individual alignment identified in the main body is observed in the case of Figures~\ref{fig:h2-cmm} and \ref{fig:h2-pmm}, which show the results of conditional MM and profile MM, respectively. The case of subset analysis shown in Figures~\ref{fig:h2-samce} and \ref{fig:h2-scamce} also reveals similar results. In particular, the case of \textcite{hankinson_when_2018} again appears problematic. The intervals are extremely wide, and performance varies sharply across subgroup conditions. Notably, the median Pearson $r$ is below 0. Similarly, \textcite{ono_contingent_2019} observe instances where the Pearson $r$ falls below 0 when subset cAMCE is being estimated. Since subset cAMCEs combine two sources of complexity, including restricted respondent subsets and subgroup-specific attribute effects, they impose a stricter test of whether synthetic data can reproduce human heterogeneity. 

\section{Appendix: Additional Stability Ratio Results}
\begin{figure}[htbp]
    \centering
    \includegraphics[width=\linewidth]{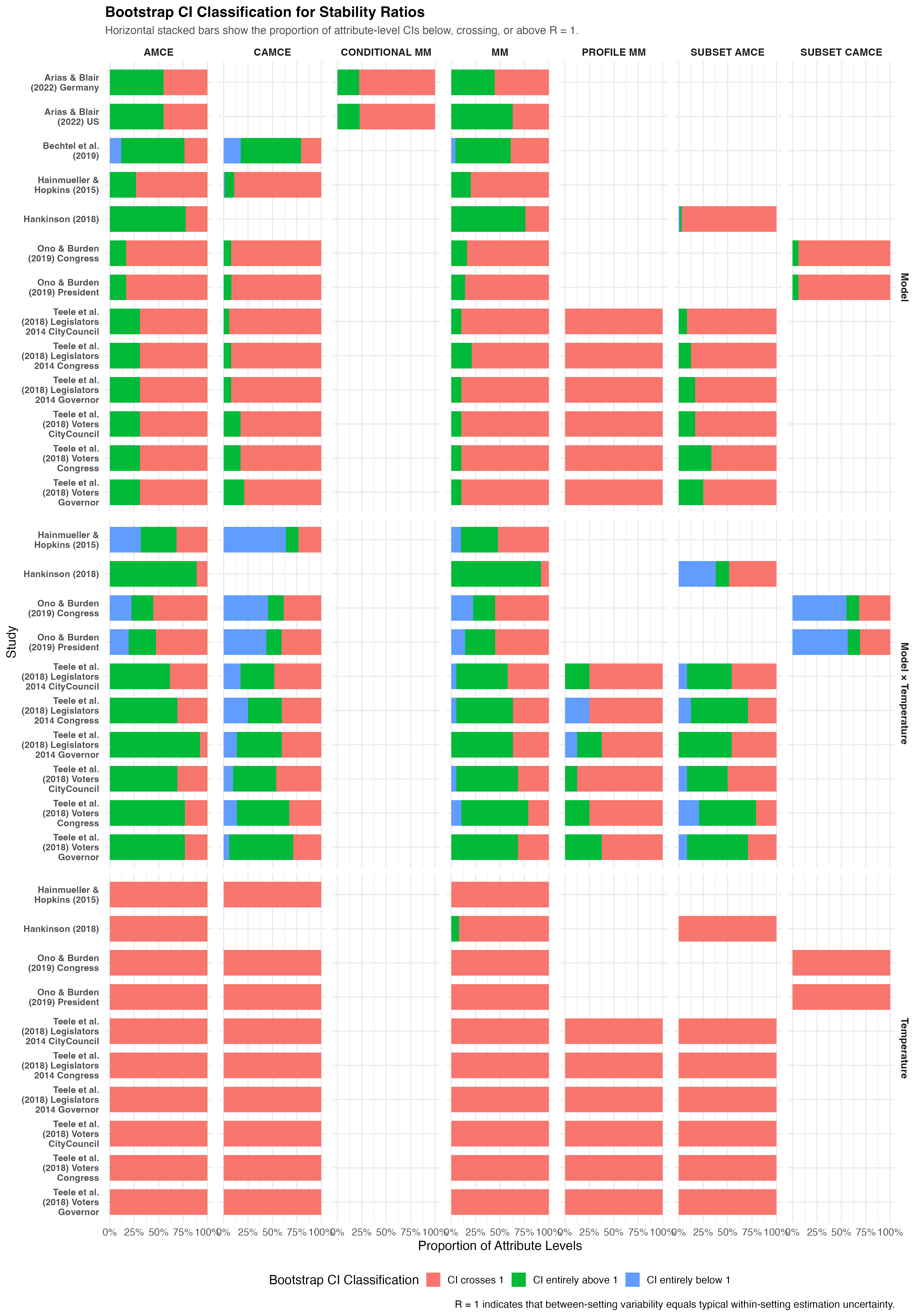}
    \caption{Bootstrap CI Classification for Stability Ratios}
    \label{fig:h3-stability-ratio-boot}
\end{figure}

\begin{figure}[htbp]
    \centering
    \includegraphics[width=\linewidth]{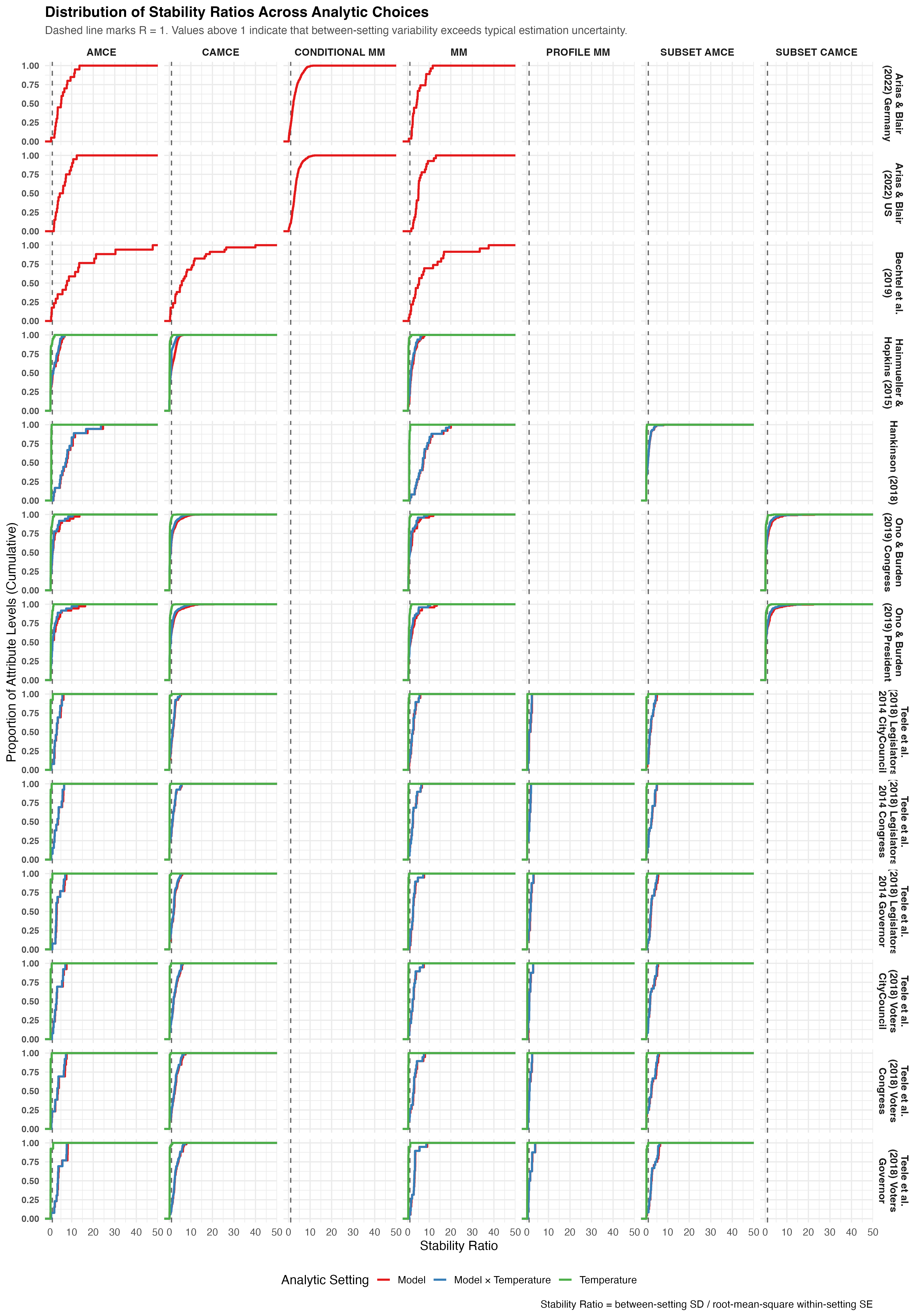}
    \caption{Distribution of Stability Ratios Across Analytic Choices}
    \label{fig:h3-stability-ratio}
\end{figure}

Figure~\ref{fig:h3-stability-ratio-boot} shows the bootstrap CI classification of the stability ratio, and Figure~\ref{fig:h3-stability-ratio} further shows the distribution of stability ratios across analytic choices. Notably, the bootstrap CI classification shows that temperature-based estimates often have CIs crossing the threshold, rather than being decisively below it. Therefore, the point estimates suggest that the temperature sensitivity is low, but the uncertainty around some temperature-based ratios remains non-negligible.

\section{Appendix: Additional Variance Analysis}
\begin{figure}[htbp]
    \centering
    \includegraphics[width=\linewidth]{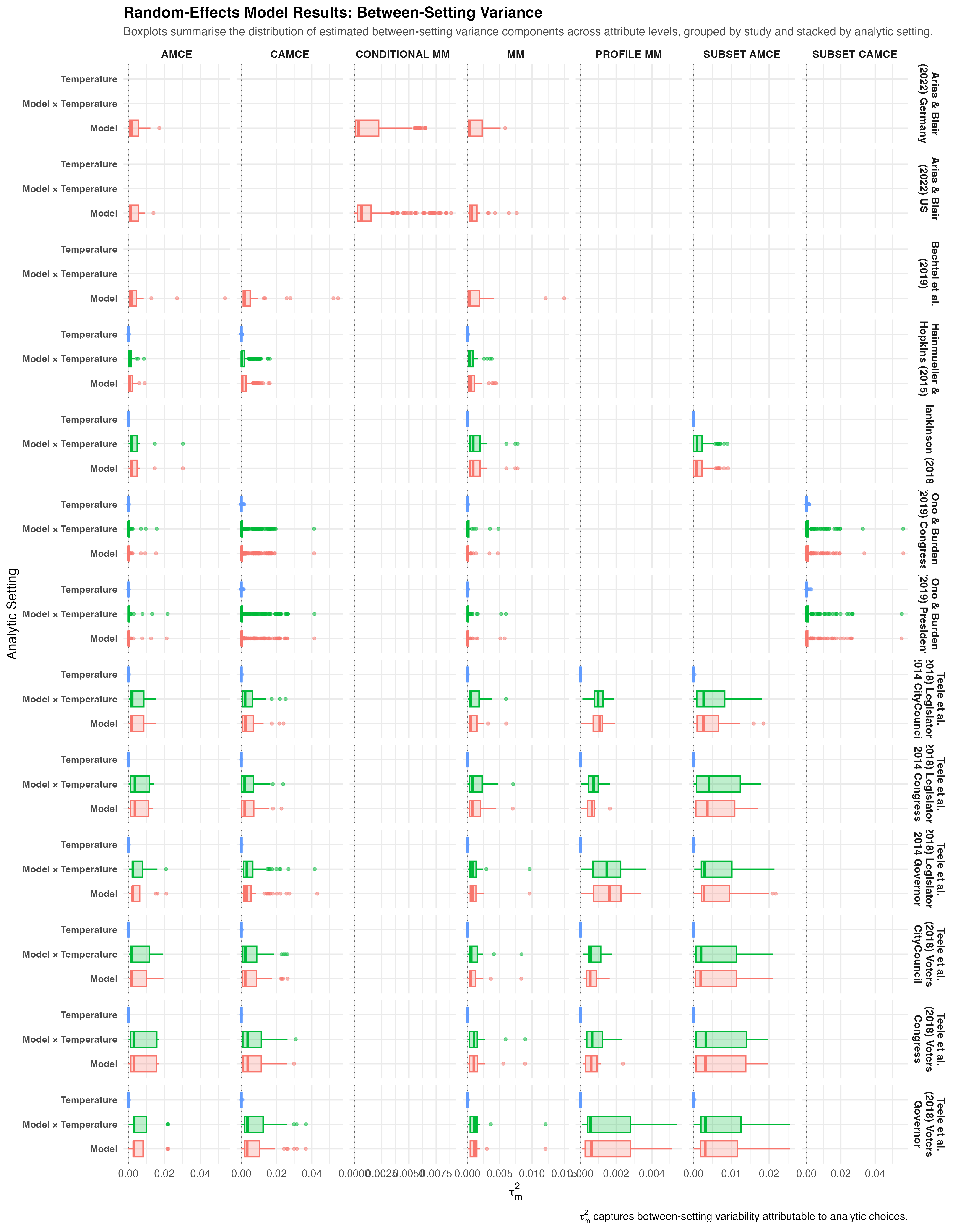}
    \caption{Random-Effects Model Results: Between-Setting Variance}
    \label{fig:h3-re}
\end{figure}

The random-effects variance estimates in Figure~\ref{fig:h3-re} provide further evidence of between-setting variability. Although the estimated $\tau_m^2$ values are often small in absolute terms, they are consistently non-zero for many attribute-level estimates. This is especially visible for model and model-by-temperature settings. When paired with the stability ratio, they together reveal that synthetic conjoint results are sensitive to analytic choices.

\begin{figure}[htbp]
    \centering
    \includegraphics[width=\linewidth]{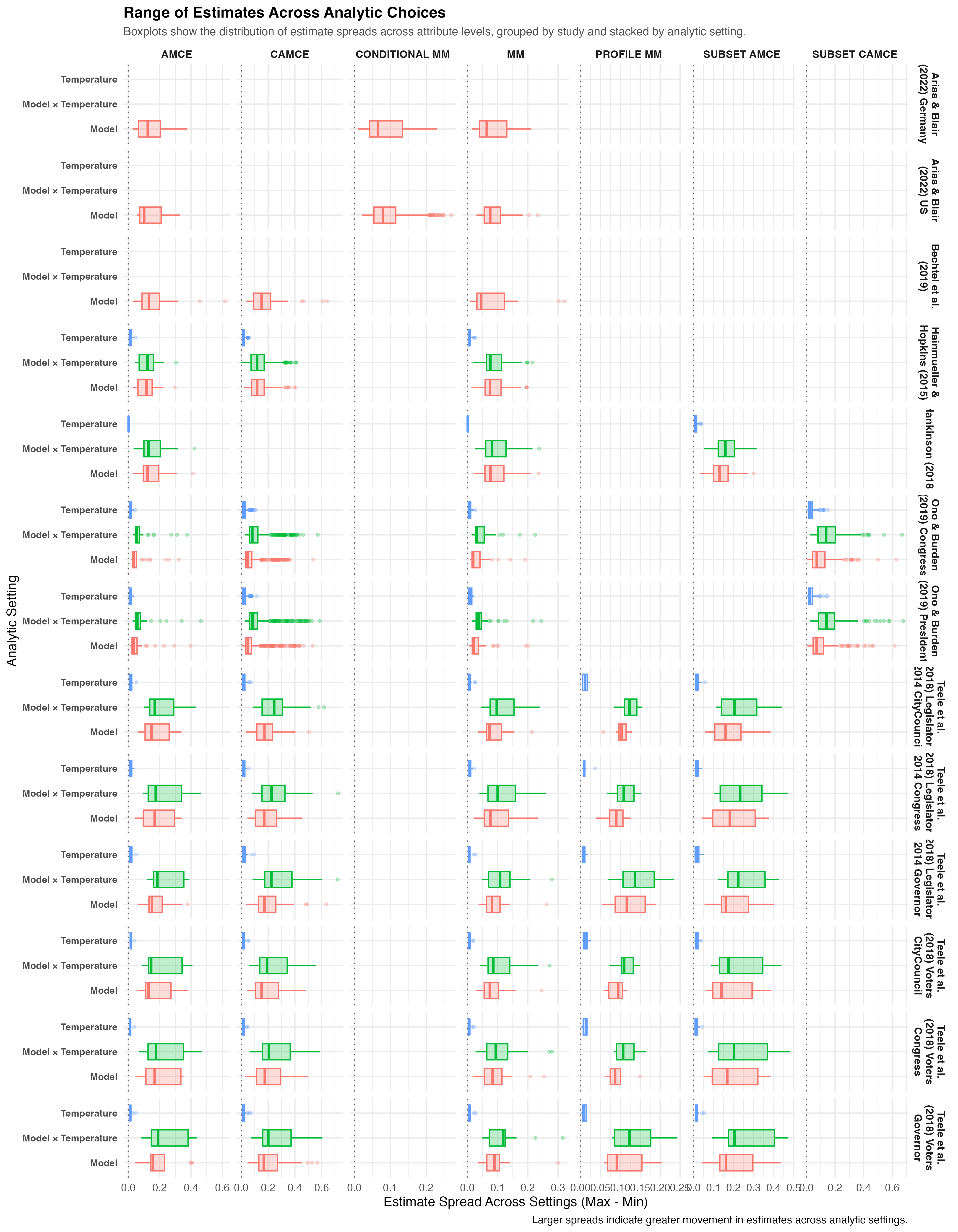}
    \caption{Range of Estimates Across Analytic Choices}
    \label{fig:h3-secondary}
\end{figure}

Finally, Figure~\ref{fig:h3-secondary} shows the range of estimates across analytic choices. Model and model-by-temperature settings generally produce wider estimate ranges than temperature-only settings.

\end{document}

%% file: preamble.tex
\usepackage[english]{babel}
\usepackage[utf8]{inputenc}
\DeclareUnicodeCharacter{FFFD}{}
\usepackage{tgheros}

\usepackage[left=2.5cm,right=2.5cm,top=2.5cm,bottom=2.5cm]{geometry}

\usepackage[justification=RaggedRight, font=small, labelfont=bf, textfont=sl, format=hang, margin={1.5cm}]{caption}
\usepackage{amsmath} 

\usepackage{newpxtext}
\usepackage{newpxmath}

\usepackage{marvosym}

\usepackage{graphicx}
\usepackage{booktabs}
\usepackage{tabularx}
\usepackage{array}
\usepackage{multirow}
\usepackage{colortbl}
\usepackage{wrapfig}
\newcolumntype{L}[1]{>{\raggedright\arraybackslash}p{#1}}
\newcolumntype{C}[1]{>{\centering\arraybackslash}p{#1}}
\newcolumntype{R}[1]{>{\raggedleft\arraybackslash}p{#1}}

\usepackage{xcolor}
\definecolor{graylight}{rgb}{0.95,0.95,0.95}
\definecolor{green2}{RGB}{0, 167, 76}
\definecolor{blue3}{HTML}{5B51EB}
\definecolor{red1}{HTML}{F30000}

\usepackage{hyperref}
\hypersetup{
    colorlinks=true,
    linkcolor=blue,
    citecolor=blue,
    urlcolor=blue,
    pdftitle={Document Title},
    pdfauthor={Your Name},
    pdfsubject={Subject},
    pdfkeywords={Keyword1, Keyword2},
    pdfproducer={LaTeX with hyperref},
    pdfcreator={XeLaTeX, Overleaf}
}
\usepackage{bookmark}

\usepackage{lineno}
\modulolinenumbers[5]

\usepackage{fancybox}
\usepackage{setspace}
\usepackage{verbatim}
\usepackage{lipsum}
\usepackage{silence}
\usepackage[hang, flushmargin]{footmisc}